\documentclass[usenatbib]{mn2e}  

\def\lapp{\ifmmode\stackrel{<}{_{\sim}}\else$\stackrel{<}{_{\sim}}$\fi}
\def\gapp{\ifmmode\stackrel{>}{_{\sim}}\else$\stackrel{<}{_{\sim}}$\fi}

\title[29 Glitches Detected at Urumqi Observatory]
     {29 Glitches Detected at Urumqi Observatory}
\author[J. P. Yuan et al.]
       {J. P. Yuan,$^{1,2}$ N. Wang,$^{1}$\thanks{E-mail: na.wang@uao.ac.cn}
 R. N. Manchester$^{3}$, Z. Y. Liu,$^{1}$\\
$^{1}$Urumqi Observatory, NAOC, 40-5 South Beijing Road, Urumqi, Xinjiang, China, 830011\\
$^{2}$Graduate University of the Chinese Academy of Sciences, 19A Yuquan road,
Beijing, China, 100049\\
$^{3}$Australia Telescope National Facility, CSIRO, PO Box 76, Epping, NSW 1710, Australia}

\date{\today}

\usepackage{psfig}   
\usepackage{lscape}   
\usepackage{booktabs}   
\usepackage{threeparttable} 

\begin{document}

\label{firstpage}  

\maketitle

\begin{abstract}
  Glitches detected in pulsar timing observations at the Nanshan radio
  telescope of Urumqi Observatory between 2002 July and 2008 December
  are presented. In total, 29 glitches were detected in 19 young
  pulsars, with this being the first detection of a glitch in 12 of
  these pulsars.  Fractional glitch amplitudes range from a few parts
  in $10^{-11}$ to $3.9\times 10^{-6}$. Three ``slow'' glitches are
  identified in PSRs J0631+1036, B1822$-$09 and B1907+10. Post-glitch
  recoveries differ greatly from pulsar to pulsar and for different
  glitches in the same pulsar. Three small glitches in PSRs B0402+61,
  B0525+21 and J1853+0545 show evidence for normal post-glitch
  recovery, but for PSRs B0144+59 and B2224+65 the spin frequency
  $\nu$ continually increases relative to the pre-glitch solution for
  hundreds of days after their small glitches. Most large glitches
  show some evidence for exponential post-glitch recovery on
  timescales of 100 -- 1000 days, but in some cases, e.g., PSR
  B1758$-$23, there is little or no recovery, that is, no detectable
  increase in $|\dot\nu|$ at the time of the glitch. Beside
  exponential recoveries, permanent increases in slowdown rate are
  seen for the two large glitches in PSRs B1800$-$21 and
  B1823$-$13. These and several other pulsars also show a linear
  increase in $\dot\nu$ following the partial exponential recovery,
  which is similar to the Vela pulsar post-glitch behaviour. However,
  the rate of increase, $\ddot\nu$, is an order of magnitude less than
  for the Vela pulsar. We present improved positions for 14 of the
  glitching pulsars. Analysis of the whole sample of known glitches
  show that fractional glitch amplitudes are correlated with
  characteristic age with a peak at about $10^5$ years, but there is a
  spread of two or three orders of magnitude at all ages. Glitch
  activity is positively correlated with spin-down rate, again with a
  wide spread of values. For some (but not all) pulsars there is a
  correlation between glitch amplitude and the duration of the
  following inter-glitch interval. In no case is there a correlation
  of glitch amplitude with the duration of the preceding inter-glitch
  interval.
\end{abstract}


\section{Introduction} \label{sec:intro} After allowing for the
accurate determination of the pulsar spin-down, pulse arrival-time
measurements display two types of irregularities in the pulsar
rotation rate: timing noise and glitches. Timing noise is a continuous
wandering of the pulsar rotation rate, on time-scales of days, months
and years. In most young pulsars, timing noise is generally very
``red'', i.e., stronger at lower frequencies \citep{cd85}, and the
second time-derivative of the pulse frequency is commonly used as an
indicator of the level of noise.

Glitches are characterized by a sudden increase in the pulsar rotation
rate, often followed by a (partial) recovery toward the pre-glitch
status. Glitches were first observed in the Vela pulsar
\citep{rm69,rd69} and it was soon realized that they could be an important
diagnostic for studying neutron-star interiors \citep{rud69,
  bpp69}. It is widely believed that these events are caused by either
sudden and irregular transfer of angular momentum from a faster
rotating interior super-fluid to the solid crust of the neutron star,
or quakes in a solid component of the star, or even both. The result
is a sudden increase in the rotational frequency $\nu$ of
the pulsar with a relative magnitude normally in the range
$10^{-10}<\Delta\nu_g/\nu<5\times10^{-6} $. Large glitches
($\Delta\nu_g/\nu > 10^{-7}$) are usually accompanied by jumps in the
spin-down rate ($\Delta\dot\nu_g/\dot\nu \sim 10^{-4} $ to $10^{-2}$).
A characteristic feature of many glitches is a relaxation after the
frequency jump, which may occur over periods of minutes to years. In
some cases, this relaxation is approximately exponential toward the
extrapolated pre-glitch state, but many different forms of relaxation
are observed, some with multiple timescales. \citet{lsg00} presented
evidence that  the degree of relaxation in different pulsars is
related to their spin-down rate $|\dot\nu|$.

Although glitches are a rather rare phenomena, an increasing number of
these events have been reported \citep{wmp+00, klgj03, js06},
permitting more comprehensive statistical studies
\citep[e.g.,][]{lsg00}. To date about 170 glitches have been observed,
in 51 of the over 1800 known pulsars\footnote{See the ATNF Pulsar
  Catalogue \citep{mhth05} Glitch table at
  www.atnf.csiro.au/research/pulsar/psrcat}. The most frequent
glitches have been detected in PSRs B0531+21 (Crab), J0537$-$6910,
B0833$-$45 (Vela), J1341$-$6220 and B1737$-$30, which have
characteristic ages in the range $\sim 10^{3}$ to $10^{5}$~ yr. PSR
J0537$-$6910 has glitched 23 times in seven years, with most of the
glitches being relatively small in $\Delta\nu_g/\nu$ terms (typically
a few parts in $10^{-7}$).  Their amplitude shows a strong correlation
with the interval to the following glitch \citep{mmw+06}. Glitches in
the Crab pulsar are even smaller, with $\Delta\nu_g/\nu$ typically
$\sim 10^{-9}$  \citep{lps93}. There is normally a rapid exponential
decay over several days
and a persistent increase in slow-down rate \citep{lps93}. The Vela
pulsar is the prototypical pulsar showing large glitches, with
$\Delta\nu_g/\nu$ generally $> 10^{-6}$ although some small and
intermediate-sized glitches have been observed. The post-glitch decay
is initially exponential, with evidence for several timescales ranging
from 1 minute to several hundred days, but subsequently linear till
the next glitch \citep{dlm07,lpgc96}. PSR B1737$-$30 is a frequently
glitching pulsar which shows a large range of glitch sizes and a high
value of the glitch activity parameter
\begin{equation}\label{eq:activity}
A_g = \frac{1}{T}\sum\frac{\Delta\nu_g}{\nu},
\end{equation}
where $T$ is the total data span \citep{ml90,sl96,zwm+08}. Significant
post-glitch decays are observed for the larger glitches.

Clearly, the glitch phenomenon in pulsars is complex. Further
observations of a larger sample of pulsars will help to quantify and
understand processes responsible for the very different glitch and
post-glitch behaviours observed both between different pulsars and in
different glitches in the same pulsar.

Nearly 280 pulsars are being monitored regularly at Urumqi Observatory in
a timing program which commenced in 2000. The aims of the
project are to make precise measurements of the position and
timing behaviour of pulsars, to provide ephemeris for observations at
radio and other wavelengths \citep[e.g.,][]{sgc+08} and to
investigate the interiors of neutron stars by monitoring
irregularities in their rotation behaviour. In this paper we describe
the glitches detected in our observations, together with updated
descriptions of the relaxations for some previously reported glitches.

\section{Observations and Analysis} \label{sec:obs}

Timing observations of pulsars using the Urumqi Observatory 25-m radio
telescope at Nanshan started in 2000 January with a dual-channel
room-temperature receiver. A dual-channel cryogenic receiver system
has been used at the central observing frequency of 1540 MHz since
2002 July. Approximately 280 pulsars are observed regularly three
times each month and results to 2008 December are reported here. The two
polarizations, each of bandwidth 320~MHz, are sent to a filter-bank
consisting of $2\times128$ channels of width 2.5~MHz. The data are
digitized to one-bit precision, and the integration time is typically
4 to 16~min \citep{wmz+01}.

Off-line data reduction was performed in the following steps. The data
for each pulsar were dedispersed and summed to produce a total
intensity profile centered at 1540~MHz. Local arrival times were
determined by correlating the data with standard pulse profiles; the
pulse times of arrival (TOAs) normally correspond to the peak of the
main pulse. TOAs at the Solar-system barycenter were obtained using
the standard timing program TEMPO\footnote{See
  http://www.atnf.csiro.au/research/pulsar/tempo} with the Jet
Propulsion Laboratories DE405 ephemeris \citep{sta98b}. A model for
the pulsar timing was fitted to these barycentric TOAs, weighted by
the inverse square of their uncertainty. 
Errors in the fitted parameters are taken to be twice
the standard errors obtained from TEMPO. The basic timing model for
the barycentric pulse phase, $\phi$, as a function of time $t$, is
\begin{equation}
   \phi(t)=\phi_{0}+\nu(t-t_{0})+\frac{1}{2}\dot{\nu}(t-t_{0})^{2}+\frac{1}{6}\ddot{\nu}(t-t_{0})^{3},
\label{eq:timing}
\end{equation}
where $\phi_{0}$ is the phase at time $t_{0}$ and $\nu$, $\dot{\nu}$,
$\ddot{\nu}$ represent the  pulse frequency, frequency derivative and
frequency second derivative respectively. Pulsar positions were derived
from long inter-glitch intervals with the residuals being ``whitened''
by fitting of higher-order frequency derivatives if necessary.

Glitches can be described as
combinations of steps in $\nu$ and $\dot\nu$, of
which parts may decay exponentially on various timescales. We assume a
glitch model for the observed phenomenon as below:
\begin{equation}
   \nu(t)=\nu_{0}(t)+\Delta\nu_{p}+\Delta\dot{\nu}_{p}t+\Delta\nu_{d}e^{-t/\tau_{d}}
 \label{eq:glitch}
\end{equation}
\begin{equation}
   \dot{\nu}(t)=\dot{\nu}_{0}(t)+\Delta\dot{\nu}_{p}+\Delta\dot{\nu}_{d}~e^{-t/\tau_{d}}
 \label{eq:glitch2}
\end{equation}
where $\Delta\nu_p$ and $\Delta\dot{\nu}_p$ are, respectively, the
permanent changes in frequency and frequency derivative relative to
the pre-glitch values and $\Delta \nu_d$ is the amplitude of the
exponential recovery with a decay time constant of
$\tau_d$. We note that for glitches the jump in frequency
  $\nu$ is positive and the jump in frequency derivative $\dot\nu$ is
  usually negative. Since $\dot{\nu}$ is negative, a negative jump
  corresponds to an increase in magnitude of the spin-down rate. The
total frequency change at the glitch is
\begin{equation}
   \Delta\nu_g= \Delta\nu_{p}+\Delta\nu_{d}.
     \label{eq:glitch3}
\end{equation}
The degree of recovery is often described by the parameter $Q$:
\begin{equation}
   Q= \Delta\nu_{d}/\Delta\nu_g.
     \label{eq:glitch4}
\end{equation}

\begin{table*}
\begin{center}
\centering
\begin{minipage}{188mm}
\caption{Parameters  for the pulsars with observed glitches}
\label{tb:position}
\begin{threeparttable}
\begin{tabular}{@{}clllllr@{.}lrrc@{}}
\hline
\multicolumn{2}{c}{Pulsar Name} & ~~~RA        &  ~~~Dec      & Epoch   & ~$P$   &\multicolumn{2}{c}{$\dot{P}$}  & Age       & ~~$B_s$\hspace{2mm} & References \\
              &                & (h ~m ~s)     & (d ~m ~s )   & MJD     & (s)   & \multicolumn{2}{c}{($10^{-15}$)}  & (kyr)  & ($10^{12}$G)     & for position  \\
\hline
J0147+5922   & B0144+59   & 01:47:44.6457(17)&  +59:22:03.293(15)& 53160   &  0.1963    &   0&256       & 12100.0   & 0.23   & this work \\
J0406+6138   & B0402+61   & 04:06:30.082(3)  &  +61:38:41.04(3)  & 53784   &  0.5945    &   5&576       &  1690.0   & 1.84   & this work \\
J0528+2200   & B0525+21   & 05:28:52.264(9)  &  +22:00:04(2)     & 53100   &  3.7455    &  40&036       &  1480.0   & 12.40  & this work \\
J0631+1036   &            & 06:31:27.524(4)  &  +10:37:02.5(3)   & 53850   &  0.2877    & 104&683       &    43.6   &  5.55  & this work \\
J1705$-$3423 &            & 17:05:42.363(3)  &$-$34:23:45.17(12) & 53830   &  0.2554    &   1&075       &  3760.0   &  0.53  & this work \\
\\
J1740$-$3015 & B1737$-$30 & 17:40:33.82(1)   &$-$30:15:43.5(2)   & 52200   &  0.6068    & 466&358       &    20.6   & 17.00  & \tnote{1} \\
J1751$-$3323 &            & 17:51:32.725(11) &$-$33:23:39.6(9)   & 53750   &  0.5482    &   8&897       &   976.0   &  2.23  & this work \\
J1801$-$2304 & B1758$-$23 & 18:01:19.829(3)  &$-$23:04:44.2(2)   & 50809   &  0.4158    & 112&880       &    58.4   &  6.93  & \tnote{2} \\
J1803$-$2137 & B1800$-$21 & 18:03:51.411(1)  &$-$21:37:07.35(1)  & 51544   &  0.1336    & 134&100       &    15.8   &  4.28  & \tnote{3} \\
J1812$-$1718 & B1809$-$173& 18:12:07.208(8)  &$-$17:18:29.5(11)  & 49612   &  1.2053    &  19&770       &  1000.0   &  4.85  & \tnote{4} \\
\\
J1818$-$1422 & B1815$-$14 & 18:18:23.7660(17)&$-$14:22:36.71(16) & 54000   &  0.2914    &   2&038       &  2270.0   &  0.78  & this work \\
J1825$-$0935 & B1822$-$09 & 18:25:30.629(6)  &$-$09:35:22.3(3)   & 53300   &  0.7689    &  52&285       &   233.0   &  6.42  & this work \\
J1826$-$1334 & B1823$-$13 & 18:26:13.175(3)  &$-$13:34:46.8(1)   & 52400   &  0.1015    &  75&061       &    21.4   &  2.79  & this work \\
J1841$-$0425 & B1838$-$04 & 18:41:05:663(1)  &$-$04:25:19.50(7)  & 53900   &  0.1861    &   6&391       &   461.0   &  1.10  & this work \\
J1853+0545   &            & 18:53:58.411(1)  &  +05:45:55.26(3)  & 53882   &  0.1264    &   0&611       &  3280.0   &  0.28  & this work \\
\\
J1902+0615   & B1900+06   & 19:02:50.277(2)  &  +06:16:33.41(6)  & 53390   &  0.6735    &   7&706       &  1380.0   &  2.31  & this work \\
J1909+1102   & B1907+10   & 19:09:48.694(2)  &  +11:02:03.35(3)  & 49912   &  0.2836    &   2&639       &  1700.0   &  0.88  & \tnote{4} \\
J1915+1009   & B1913+10   & 19:15:29.984(2)  &  +10:09:43.67(4)  & 53300   &  0.4045    &  15&251       &   420.0   &  2.51  & this work \\
J2225+6535   & B2224+65   & 22:25:52.721(7)  &  +65:35:35.58(4)  & 53880   &  0.6825    &   9&659       &  1130.0   &  2.58  & this work \\

\hline
\end{tabular}
   \begin{tablenotes}
     \item[] 1 \citet[]{fgm+97}; 2. \citet[]{fkv93}; 3. \citet[]{bck06}; 4. \citet[]{hfs+04}.
    \end{tablenotes}
\end{threeparttable}
\end{minipage}
\end{center}
\end{table*}

Because of the decaying component, the instantaneous
 change in $\dot{\nu}$ at the glitch differs from $\Delta\dot{\nu}_p$
\begin{equation}
    \Delta\dot{\nu}_g = \Delta\dot{\nu}_{p} + \Delta\dot{\nu}_{d} =
             \Delta\dot{\nu}_{p} - Q\Delta\nu_g/\tau_{d}
     \label{eq:glitch5}
\end{equation}
Solutions before and after the glitch are used to estimate the steps
in frequency and its derivative.  If the gap between observations
around the glitch is not too large, the glitch epoch can be estimated
more accurately by requiring a phase-connected solution over the gap
in the TEMPO fit. However if the gap is too large, rotations can be
missed and the glitch epoch is estimated as the mid-point of the gap
between observations around the glitch.

\begin{table*}
\begin{center}
 \centering
 \begin{minipage}{160mm}
 \caption{Pre- and post-glitch timing solutions}
 \label{tb:frequency}
  \begin{tabular}{@{}lllrrlrr@{}}
  \hline
 Pulsar Name &$\nu$     & $\dot{\nu}$          & $\ddot{\nu}$~~     & Epoch   & MJD Range & No. of & RMS          \\
             &(s$^{-1}$) & ($10^{-12}$~s$^{-2})$ & $(10^{-24}$~s$^{-3})$ & (MJD)  &           & TOAs   &  ($\mu$s) \\
 \hline
   B0144+59  &  5.093689334287(4) & $-$0.0066633(2)     &  $-$             & 53160.0   & 52486$-$53662 &  90   &  101\\
\vspace{2.1mm} 
             &  5.093688844679(4) & $-$0.0066619(2)     &  $-$             & 54011.0   & 53686$-$54831 & 107   &  145\\
   B0402+61  &  1.68187172559(1)  & $-$0.015757(2)      & $-$              & 52745.0   & 52469$-$53022 &  57   &  395\\
\vspace{2.1mm} 
             &  1.681870311647(2) & $-$0.01576474(8)    & $-$              & 53784.0   & 53050$-$54830 & 189   &  541\\
   B0525+21  &  0.266984828978(2) & $-$0.0028537(3)     &  $-$             & 51864.0   & 51546$-$52271 &  85   &  575\\
             &  0.266984598507(1) & $-$0.00285481(3)    &  $-$             & 52800.0   & 52287$-$53975 &  249  &  834\\
\vspace{2.1mm} 
             &  0.266984253314(2) & $-$0.0028550(1)     & $-$              & 54200.0   & 53982$-$54831 &  182  &  817\\
J0631+1036   &  3.47486317481(6)  & $-$1.263638(6)      & 30(2)            & 52657.0   & 52473$-$52842 &  40   &  274 \\
             &  3.47482575938(6)  & $-$1.265058(9)      & 106(1)           & 53000.0   & 52853$-$53219 &  37   &  685 \\
             &  3.474732970746(8) & $-$1.263032(1)      & 5.47(8)          & 53850.0   & 53238$-$54100 & 175   & 1454 \\
             &  3.47467842199(3)  & $-$1.263630(2)      & 15.7(6)          & 54350.0   & 54129$-$54630 & 106   & 5003 \\
\vspace{2.1mm} 
             &  3.4746348982(1)   & $-$1.26374(2)       & 166(12)          & 54750.0   & 54634$-$54831 &  57   & 315  \\
J1705$-$3423 &  3.915021810541(8) & $-$0.0164948(2)     & $-$              & 53800.0   & 52486$-$54384 &  222   & 670\\
\vspace{2.1mm} 
             &  3.9150208150(1)   & $-$0.016495(9)      & $-$              & 54500.0   & 54394$-$54832 &   44   & 590\\
 B1737$-$30  &  1.647839443329(3) & $-$1.2659499(3)     & 13.16(9)         & 54000.0   & 53101$-$54394 & 115   & 1962 \\
             &  1.64778155382(3)  & $-$1.26561(1)       & 59(3)            & 54530.0   & 54463$-$54685 &  19   &  314 \\
\vspace{2.1mm} 
             &  1.6477542263(1)   & $-$1.26557(3)       & $-$              & 54780.0   & 54695$-$54808 &   7   &  338 \\
J1751$-$3323 &  1.82406604540(3)  & $-$0.029722(4)      & $-$              & 52758.0   & 52496$-$52994 &  35   & 1121\\
             &  1.82406350461(2)  & $-$0.0297075(4)     & $-$0.58(4)       & 53750.0   & 53007$-$54384 &  75   & 3667\\
\vspace{2.1mm} 
             &  1.8240615852(2)   & $-$0.02938(2)       & $-$              & 54500.0   & 54463$-$54714 &  18   & 1118\\
 B1758$-$23  &  2.40490661705(4)  & $-$0.65316(2)      & 3(3)             & 52900.0   & 52494$-$53292 &  67   & 3079   \\
\vspace{2.1mm} 
             &  2.40484572830(1)  & $-$0.6531075(6)    & 1.12(4)          & 54000.0   & 53327$-$54831 & 116   & 2604   \\
 B1800$-$21  &  7.48196609098(1)  & $-$7.4862079(4)     & 238.57(3)        & 52800.0   & 51916$-$53429 & 165   & 7091  \\
\vspace{2.1mm} 
             &  7.481282565107(9) & $-$7.5200390(9)     & 207.93(4)        & 53900.0   & 53458$-$54832 & 108   & 12533 \\
 B1809$-$173 & 0.82961975150(3)  & $-$0.013100(4)      &  $-$             & 52800.0   & 52529$-$53105 &  32   & 1360  \\
             &  0.82961862949(2)  & $-$0.0131283(4)     & 0.25(5)          & 53800.0   & 53119$-$54351 &  79   & 1013  \\
\vspace{2.1mm} 
             &  0.82961772349(6)  & $-$0.01313(1)       &  $-$             & 54600.0   & 54382$-$54764 &  17   & 1214 \\
 B1815$-$14  &  3.43066049095(2)  & $-$0.024011(3)      &  $-$             & 51822.0   & 51512$-$52048 &  69   &  558 \\
\vspace{2.1mm} 
             &  3.430655769877(4) & $-$0.0239594(2)     & $-$0.446(3)      & 54100.0   & 52081$-$54831 & 226   & 2400 \\
 B1822$-$09  &  1.300387566396(6) & $-$0.0886064(2)     & 4.34(4)          & 53280.0   & 52811$-$53716 & 83    & 1362 \\
             &  1.30038260224(2)  & $-$0.088638(3)      & $-$22.0(6)       & 53930.0   & 53723$-$54094 & 45    & 1203 \\
\vspace{2.1mm} 
             &  1.30038021608(2)  & $-$0.088780(2)      & $-$2.8(1)        & 54262.0   & 54129$-$54831 &  54   & 1543  \\
 B1823$-$13  &  9.85460718889(1)  & $-$7.2729146(3)     & 105.74(2)        & 52400.0   & 51574$-$53199 & 132   & 3407 \\
             &  9.85394781571(6)  & $-$7.265061(9)      & 66(1)            & 53450.0   & 53254$-$53730 &  32   &  389 \\
\vspace{2.1mm} 
             &  9.85349875132(2)  & $-$7.3062994(9)     & 280.88(9)        & 54200.0   & 53739$-$54518 &  90   & 14061 \\
 B1838$-$04  &  5.37205530549(3)  & $-$0.184289(2)      & 1.5(2)           & 53044.0   & 52535$-$53387 &  57   &  364  \\
\vspace{2.1mm}
             &  5.37204477631(1)  & $-$0.1844913(5)     & $-$1.56(4)       & 53900.0   & 53429$-$54662 &  99   &  556  \\
J1853+0545   &  7.91137412600(2)  & $-$0.038303(1)      & $-$              & 53100.0   & 52493$-$53434 &  63   &  446\\
\vspace{2.1mm} 
             &  7.91137154739(1)  & $-$0.0383351(1)     & $-$0.19(5)       & 53882.0   & 53458$-$54830 & 113   &  549\\
   B1900+06  &  1.484775165625(2) & $-$0.01700231(8)    & $-$              & 53390.0   & 52473$-$54240 &  129  &  340\\
\vspace{2.1mm} 
             &  1.48477368242(2)  & $-$0.017003(1)      & $-$              & 54400.0   & 54262$-$54821 &   39  &  306 \\
 B1907+10    &  3.52558388053(5)  & $-$0.03286(3)       & $-$              & 52578.0   & 52470$-$52686 &  18   &  161\\
             &  3.525581831083(6) & $-$0.0328630(5)     & $-$              & 53300.0   & 52705$-$53689 &  71   &  391\\
\vspace{2.1mm} 
             &  3.52557928013(1)  & $-$0.0328110(3)     & $-$2.02(4)       & 54200.0   & 53701$-$54821 &  88   &  350\\
 B1913+10    &  2.471906137401(5) & $-$0.0931843(1)     & $-$0.071(9)      & 53308.0   & 52473$-$54159 & 118   &  311 \\
\vspace{2.1mm} 
             &  2.47189777002(2)  & $-$0.093219(1)      & $-$              & 54348.0   & 54180$-$54831 &  61   &  306 \\
 B2224+65    &  1.465111203809(5) & $-$0.0207375(4)     & $-$0.268(9)      & 53880.0   & 52487$-$54265 & 220   &  922 \\

             &  1.46511023680(2)  & $-$0.020737(2)      & $-$              & 54420.0   & 54275$-$54831 &  71   &  484 \\
 \hline
\end{tabular}
\end{minipage}
\end{center}
\end{table*}

\section{Results} \label{sec:res} 

We present the results of glitch analysis using Nanshan timing data
from July 2002 to Dec 2008. Table~\ref{tb:position} lists the
parameters for the 19 pulsars for which glitches were observed. The
first two columns give the pulsar names based on J2000 and B1950
coordinates and the third and fourth columns are their J2000
positions.  The next four columns give the pulsar period and period
derivative, characteristic age $\tau_c = P/(2\dot P)$ and surface
dipole magnetic field $B_s = 3.2\times 10^{19} \sqrt{P\dot P}$~G. Most
of the positions quoted in Table~\ref{tb:position} result from our
work except for five pulsars where the referenced positions are more
accurate.  Estimated uncertainties in the last quoted digit,
  given in brackets, are 2$\sigma$ from the TEMPO fit in this and
  following tables.  References for the position are given in the
  final column.  The data spans used to determine the positions are
  the largest available which are not too strongly affected by
  post-glitch recovery.  Typical spans are 3 -- 5~yr, with a minimum
  of 2.3 yr and a maximum of 7.5 yr.

 As mentioned in Section~\ref{sec:obs} the data sets used for
  position fitting are whitened by fitting for higher order
  derivatives of frequency. In most cases, five or less derivatives
  are sufficient to whiten the data, giving rms residuals of 100 --
  700 $\mu$s. For PSR J1751-3323, nine derivatives were required and
  for PSR B1815-14, seven derivatives. \citet{zhw+05} investigated the
  effect of timing noise on the determination of pulsar positions,
  showing that an uncertainty of twice the formal TEMPO error is generally
  realistic for data spans of the order of 5 yr.

In total, 29 glitches were detected in these pulsars. The
rotational parameters from independent fits to the pre-glitch and
post-glitch data are given in Table~\ref{tb:frequency}. Except for
short sections of data, the fits include a $\ddot{\nu}$ term, which
is generally dominated by timing noise or recovery from a previous
glitch.

Glitch parameters are given in Table~\ref{tb:glitch}, in which the
second column gives the glitch epoch, and the next two columns give the
fractional changes in $\nu$ and $\dot{\nu}$ at the glitch determined
by extrapolating the pre- and post-glitch solutions to the glitch
epoch. Quoted uncertainties include a contribution from the
uncertainty in glitch epoch. Results of directly fitting for the
glitch parameters in TEMPO are given in the remaining columns. Decay
terms were only fitted when significant decay was observed.

The rotational behaviour of these pulsars are shown in
Fig.~\ref{Fig:0147n} to \ref{Fig:2225n}. Values of $\nu$ and
$\dot{\nu}$ were obtained from independent fits of
Equation~\ref{eq:timing} (omitting the $\ddot\nu$ term) to data spans
ranging from 50~d to 200~d. The residual frequencies have been
obtained at the various epochs by the subtraction of the pre-glitch
models. We discuss the results for each pulsar in the following
Sections.

\subsection{PSR B0144+59}\label{B0144}
\begin{figure}
\centerline
{
 \hspace{0mm}\psfig{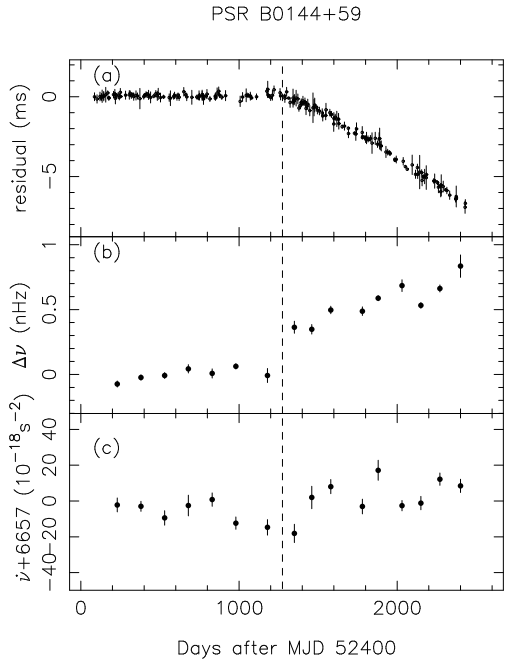}
}
\caption{The 2005 December glitch of PSR B0144+59: (a) timing
  residuals relative to the pre-glitch model.  (b) variations of the frequency
  residual $\Delta\nu$ relative to the pre-glitch solution. (c) the
  variations of $\dot\nu$.  The dashed vertical line indicates the
  epoch of the glitch.}
\label{Fig:0147n}
\end{figure}

This pulsar has a period of 196~ms and a small period derivative
implying a relatively large characteristic age of 12.1~Myr. Normally
it is a very stable pulsar and has no previously detected
glitch. Despite this, we detected a very small glitch at MJD $\sim$
53682, with a fractional frequency step of just
$0.055\times10^{-9}$. There are only four measured glitches
  with fractional size smaller than this. Also, PSR B0144+59 is the
  second oldest pulsar for which a glitch has been observed, with the
  oldest being the millisecond pulsar PSR B1821$-$24 which has a
  characteristic age of 29.9 Myr.  Fig.~\ref{Fig:0147n} shows that
after the glitch there is continuous increase of the frequency
relative to the extrapolated pre-glitch (i.e., positive
$\Delta\dot\nu_p$ or decrease in $|\dot\nu|$) that lasts for at least
600~d. This is unusual and opposite to the normal post-glitch
decay. It is possible that this tiny glitch in PSR B0144+59
is just a fluctuation in the timing noise, but it does appear distinct
in the $\Delta\nu$ plot. Similar decreases in $|\dot\nu|$
  following a glitch were observed for PSR B1951+32 \citep{fbw90} and
  in the RRAT PSR J1819$-$1458 \citep{lmk+09}. These may imply an
  increase in the moment of inertia or a decrease in the spin-down
  torque at the time of the glitch. For the PSR B0144+59 position fit
  given in Table~\ref{tb:position}, whitening of the residuals required two
  frequency derivatives.

\subsection{PSR B0402+61}

\begin{figure}
\centerline
{
 \hspace{2mm}\psfig{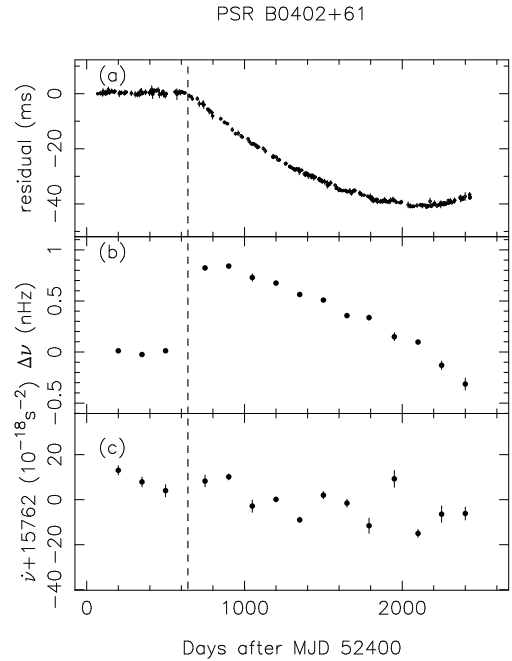}
}
\caption{The glitch of PSR B0402+61: (a) timing residuals relative to
the pre-glitch model, (b) variations of frequency residual $\Delta\nu$
 relative to the pre-glitch solution, (c) the variations of $\dot{\nu}$.}
\label{Fig:0406n}
\end{figure}

This pulsar is also relatively old with $\tau_c \sim
1.7$~Myr. Fig.~\ref{Fig:0406n} shows a small glitch at MJD $\sim$
53041 (2004 February) with $\Delta\nu_g/\nu\sim0.6\times10^{-9}$.
In contrast to PSR B0144+59, there is an almost linear decrease  of
$\nu$ relative to the pre-glitch value that lasts for at least 1000 days
and overshoots the pre-glitch extrapolation. This is similar to the behaviour
seen in the Crab pulsar \citep{lps93}.
The negative $\Delta\dot\nu_p$ is seen in  Fig.~\ref{Fig:0406n} (b)
and is listed in Table~\ref{tb:glitch}.

\subsection{PSR B0525+21}

\begin{figure}
\centerline
{
 \hspace{0mm}\psfig{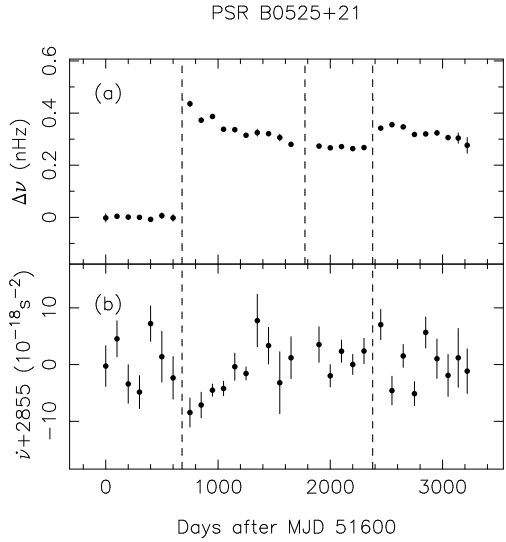}
}
\caption{Glitches of PSR B0525+21: (a) variations of frequency
residual $\Delta\nu$ relative to the pre-glitch solution, and (b) the
frequency first derivative $\dot{\nu}$. The long dashed lines are at
the epochs of well-defined glitches; the short dashed line is at the
epoch of the small glitch claimed by \citet{js06}. }
\label{Fig:0528n}
\end{figure}

This very long-period pulsar has had three small glitches
reported. The first at MJD 42057 (January 1974) had a magnitude of
$\Delta\nu_g/\nu=1.2\times10^{-9}$ and an exponential decay with
$Q=0.5$ and timescale 150 days was measured
\citep{dow82,sl96}. \citet{js06} reported a second glitch of similar
magnitude at MJD 52284 and suggest that a very small glitch with
$\Delta\nu_g/\nu=0.17\times10^{-9}$ may have occurred at MJD
53375. Our data, shown in Fig.~\ref{Fig:0528n}, confirm the glitch at
MJD 52284 (our central date is MJD 52280 but the dates are consistent
within the uncertainties). The measured $Q$ and $\tau_d$ values are
similar to those for the 1974 glitch. Our data also show another small
glitch of magnitude $\Delta\nu_g/\nu=0.43\times10^{-9}$ at MJD
53980. However, Fig.~\ref{Fig:0528n} shows no evidence for the small
glitch (of amplitude $\Delta\nu \sim 0.05$~nHz) at MJD 53375 claimed
by \citet{js06}. Although this epoch coincides with a small gap in our
data, a limit of $\sim 0.01$~nHz can be placed on the amplitude of any
glitch at this time. Since \citet{js06} did not show time or frequency
residuals for this pulsar, it is not possible to comment on what led
them to believe that there was a glitch at this time. We note that it
is the smallest of the three glitches that they reported for this
pulsar.

\subsection{PSR J0631+1036}\label{J0631}
\begin{figure}
\centerline
{
 \hspace{2mm}\psfig{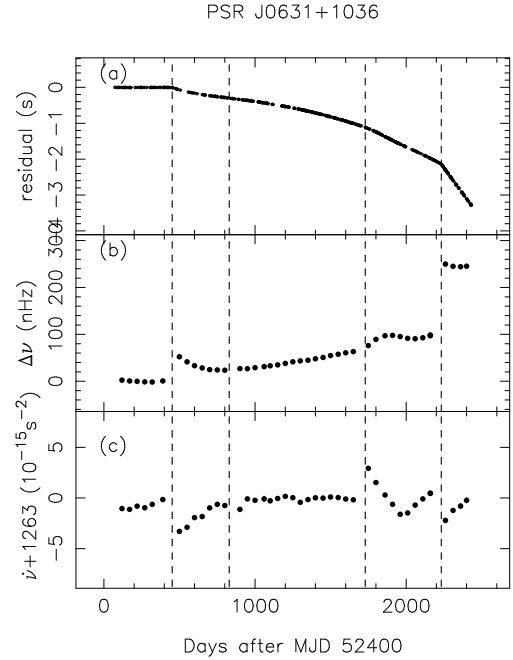}
}
\caption{Glitches of PSR J0631+1036: (a) timing residuals
relative to the pre-glitch model, (b) variations of frequency residual $\Delta\nu$
relative to the pre-glitch solution, and (c) the frequency first
derivative $\dot{\nu}$. }
\label{Fig:0631n}
\end{figure}

\begin{figure}
\centerline
{
 \hspace{0mm}\psfig{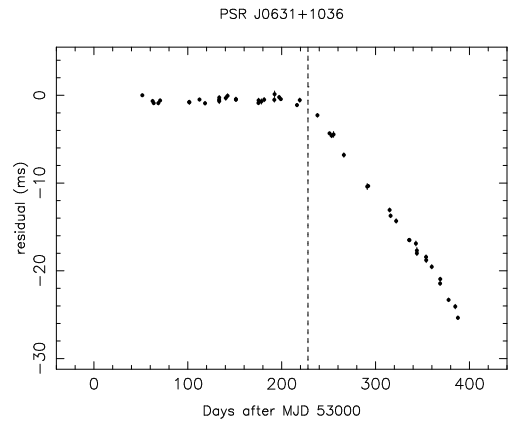}
}
\caption{Timing residuals around the time of the small glitch in PSR J0631+1036.}
\label{Fig:0631r}
\end{figure}

PSR J0631+1036, discovered by \citet{zcwl96}, has a period of 288~ms
and a very large period derivative of $105\times10^{-15}$ implying a
small characteristic age $\sim$ 44~kyr. As shown in
Fig.~\ref{Fig:0631n}, three normal glitches and one unusual event
which may be a slow glitch were detected in our data. The first glitch
at MJD 52852 has an amplitude of $\Delta\nu_g/\nu\sim19\times10^{-9}$
and shows a clear relaxation which is well described by an exponential
decay model with $\tau_{\rm d}\sim120$~d and $Q\sim0.62$
(Table~\ref{tb:glitch}).  A small second glitch was detected around
MJD $\sim$53229. Although it is not so obvious in
Fig.~\ref{Fig:0631n}, timing residuals around the time of the glitch
given in Fig.~\ref{Fig:0631r} confirm its reality. Its relative
magnitude is $\sim1.6\times10^{-9}$.

Fig.~\ref{Fig:0631n} shows another event starting at MJD $\sim$ 54129 which
may be related to the slow glitches detected in PSR B1822$-$09
\citep{sha98,zww+04,sha07}. At MJD $\sim$ 54129 there is a sudden increase in
$\dot\nu$ (decrease in $|\dot\nu|$) leading to a gradual increase in
$\nu$ of $\sim 23$~nHz over about 130~d. During this time, $\dot\nu$
steadily decreased, over-shooting the pre-event value and returning
$\nu$ to the extrapolation of its behaviour before the
event. Table~\ref{tb:glitch} lists the maximum $\Delta\nu_g/\nu$ of
$\sim 6.6\times10^{-9}$. This behaviour is similar to that of the slow
glitches observed in PSR B1822$-$09 \citep{sha07} except that the rise
in $\dot\nu$ appears more abrupt and no overshoot is observed for PSR
B1822$-$09, leading to a more permanent increase in $\nu$ relative to
the pre-glitch solution.

A fourth glitch occurred in June 2008, with an amplitude of
$\Delta\nu_g/\nu \sim 44\times10^{-9}$. Although the data span after
this event is short, it can be seen from the Fig.~\ref{Fig:0631n} that
there is a jump in $|\dot\nu| \sim 3 \times10^{-15}~$s$^{-2}$ at the
time of the glitch followed by some recovery. It is too early to say
if the recovery is exponential, similar to that for the glitch at MJD 52852.

\subsection{PSR J1705$-$3423}
\begin{figure}
\centerline
{
 \hspace{0mm}\psfig{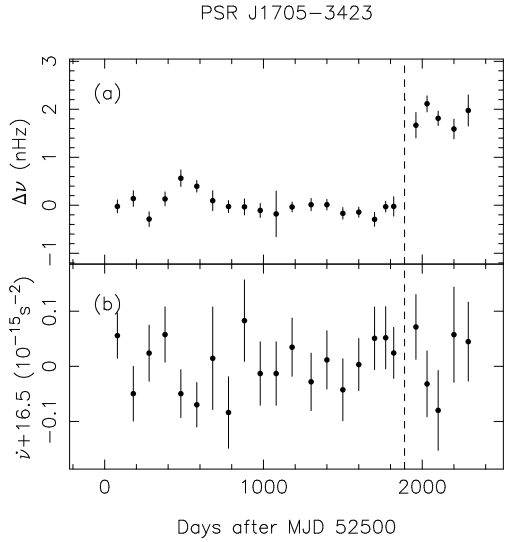}
}
\caption{Timing behaviour of PSR B1705$-$3423: (a) variations of
  rotational frequency $\Delta\nu$ relative to the pre-glitch value,
  and (b) variations of the frequency first derivative $\dot{\nu}$.}
\label{Fig:1705n}
\end{figure}

This pulsar, discovered in the Parkes Southern Pulsar Survey by
\citet{mld+96}, has a period of 0.255 s and period derivative of $1.07
\times10^{-15}$.  A small glitch was detected in Oct 2007. A plot of
the time variation of $\Delta\nu$ is given in
Fig.~\ref{Fig:1705n}a. The dominant effect is a small change in $\nu$,
with a magnitude of $\sim~2.0 \times10^{-9}$~Hz.  Fig.~\ref{Fig:1705n}b
shows no obvious change in $\dot\nu$.

\subsection{PSR B1737$-$30}
\begin{figure}
\centerline
{
 \hspace{0mm}\psfig{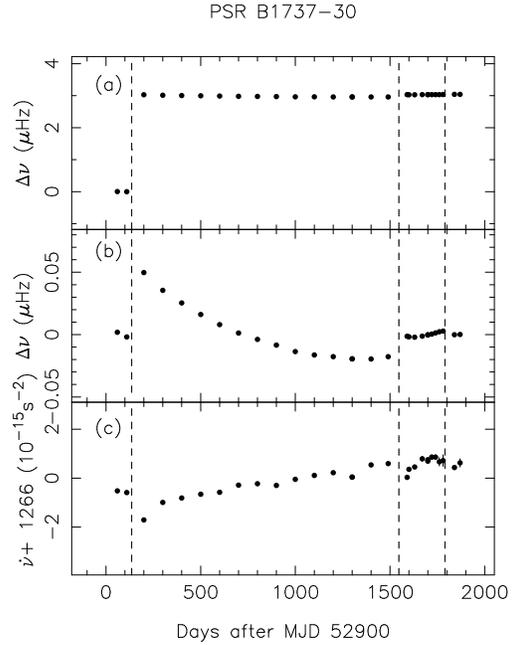}
}
\caption{Glitches of PSR B1737$-$30: (a) variations of rotational
  frequency $\Delta\nu$ relative to the pre-glitch value, (b) an
  expanded plot of $\Delta\nu$ where the mean post-glitch value has
  been subtracted  from the post-glitch data,  and (c) variations of 
the frequency  first derivative $\dot{\nu}$.}
\label{Fig:1740-3015n}
\end{figure}

PSR B1737$-$30 is a young radio pulsar which has exhibited 20 glitches
in 20 years, making it one of the most frequently glitching pulsars
known \citep{ml90,sl96,klgj03,zwm+08}. The relative glitch amplitudes
range widely from a few parts in $10^{-9}$ to nearly $2\times
10^{-6}$. Significant increases in $|\dot\nu|$ at the time of the
glitch are observed, at least for the larger glitches, and these tend
to decay linearly with time. Fig.~\ref{Fig:1740-3015n} shows the last
(giant) glitch observed by \citet{zwm+08} and two further smaller
glitches at MJD $\sim$ 54447 and $\sim$ 54694. Following the large
glitch there is a short-term ($\sim$ 100 d) exponential recovery in
$\dot{\nu}$ followed by an approximately linear increase in
$\dot{\nu}$ until the next glitch. This behaviour is very similar to
that observed in the Vela pulsar \citep{lpgc96} as will be discussed
further in \S\ref{sec:dis}. Following this large
glitch, no further glitches were observed for 1400 days, an unusually
long ``quiet'' period for this pulsar. A small glitch with
$\Delta\nu_g/\nu\sim 41\times10^{-9}$ was detected at MJD $\sim$
54447. Then, about 280~d later, we observed another small glitch with
$\Delta\nu_g/\nu\sim 2.4\times10^{-9}$. This brings the mean glitch
rate back close to one per year.

\subsection{PSR J1751$-$3323}
\begin{figure}
\centerline
{
 \hspace{2mm}\psfig{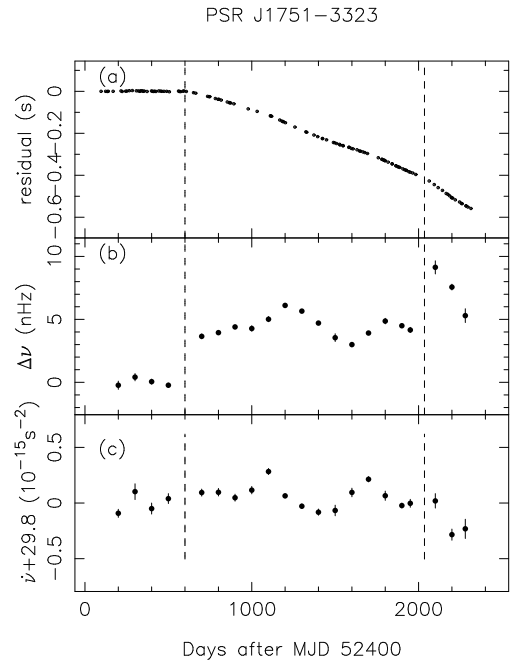}
}
\caption{The glitches of PSR J1751$-$3323: (a) timing residuals, (b)
  frequency residuals $\Delta\nu$ relative to the pre-glitch solution,
  and (c) the variations of $\dot{\nu}$.}
 \label{Fig:1751n}
\end{figure}

This pulsar was discovered in a Parkes multi-beam survey by
\citet{kbm+03}. In addition to its noisy timing behaviour, two
small glitches were detected around MJD 53004 and 54435 with
$\Delta\nu_g/\nu\sim 2\times10^{-9}$ and
$\Delta\nu_g/\nu\sim 3\times10^{-9}$ respectively.
Fig.~\ref{Fig:1751n} shows that no significant
changes in $\dot\nu$ can be associated with these glitches.

\begin{landscape}
\begin{table}
\begin{center}
\caption{The glitch parameters.}
\begin{threeparttable}
\label{tb:glitch}
\begin{tabular}{@{}llllcccccccc@{}}
\hline
            &              &          \multicolumn{2}{c}{Extrapolated}    &  & \multicolumn{6}{c} { Fitted }      \\
\cmidrule{3-4}
\cmidrule{6-12}
Pulsar  & Epoch &$\Delta\nu_g/\nu$& $\Delta\dot\nu_g/\dot{\nu}$ &  & $\Delta\nu_g/\nu$ & $\Delta\dot\nu_g/\dot{\nu}$ & $\Delta\dot\nu_p$ & Q & $\tau_{d}$ & res & Data Range\\
 Name       &   MJD &  $(10^{-9})$  & $(10^{-3}) $                 &  &  $ (10^{-9}) $  &   $(10^{-3}) $                 & ($10^{-15}$~s$^{-2}$) &  & (d)     & ($\mu$s) &     \\
 \hline
\vspace{2.1mm} 
 B0144+59   & 53682(15)*    &  0.056(3)  &  $-$0.21(5)  &   & 0.055(8)        & $-$0.22(8)  & 0.0015(8)  & $-$      & $-$      & 128    & 52486$-$54831  \\
\vspace{2.1mm} 
 B0402+61   & 53041(6)*      &  0.62(4)   &  0.5(1)      &   & 0.60(4)         & 0.47(16)   & $-$0.007(2)  &  $-$     &   $-$    & 407    & 52469$-$54517  \\
 B0525+21   & 52280(4)*        &  1.53(5)   &  0.9(1)      &   & 1.6(2)          & 1.1(1)     &   & 0.44(5)  & 650(50)  & 620    & 51547$-$53975  \\
\vspace{2.1mm} 
            & 53980(12)*       &  0.5(1)    &  0.13(12)    &   & 0.43(6)         & 0.24(18)   &   & $-$      & $-$      & 769    & 53387$-$ 54517 \\
J0631+1036  & 52852.0(2)*        &  17.6(9)   &  3.3(2)    &   & 19.1(6)           & 3.1(6)     &   & 0.62(5)    & 120(20)  & 518   & 52606$-$53143 \\
            & 53229(10)       &  1.5(4)    & $-$0.08(5)   &   & 1.6(7)          & 0.3(4)     &   & $-$      & $-$      & 249      & 53051$-$53387 \\
            & 54129--54260    &  6.6(4)    & $-$2.5(2)    &   & $-$             & $-$        &   & $-$      & $-$      &        &     \\
\vspace{2.1mm} 
            & 54632.41(14)*    & 44(2)      &  5.7(5)      &   & 44(1)           &  4(2)      &   & 0.13(2)  &  40(15)  &  257   & 54437--54750  \\

\vspace{2.1mm} 
J1705$-$3423  & 54384(10)*        &   0.51(4)   & $-$0.1(5)     &   & 0.52(5)         & 0.0(6)  &   &  $-$    & $-$       &  660    & 52486$-$54832 \\
 B1737$-$30 & 54447.41(4)*       &  42(3)     &  0.2(4)      &   & 41.0(2)         & 0.7(1)  &   & $-$      & $-$      &  1811  & 53101$-$54696 \\
\vspace{2.1mm} 
            & 54694(3)*        &  2.2(4)    &  0.61(5)     &   & 2.4(3)          & 0.86(6)    &   &  $-$     & $-$      &  299   & 54463$-$54807  \\
J1751$-$3323 & 53004(6)*       &  2.3(3)    & $-$1.4(6)    &   &  2.0(2)         & $-$2.1(4)  &   & $-$      & $-$      & 883   & 52496$-$53433 \\
\vspace{2.1mm} 
             & 54435(8)*      &  3(1)      & $-$1(3)      &   &  3.0(4)         & 1(3)       &   & $-$      & $-$      & 3683   & 53051$-$54620 \\
\vspace{2.1mm} 
  B1758$-$23 & 53309(18)       &  493.2(4)  &  0.26(5)     &   & 494(1)          &  0.19(3)   &   & 0.009(2) & 1000(100)& 2896   & 52494$-$54606 \\
\vspace{2.1mm} 
  B1800$-$21 & 53444(16)       &  3910(12)  &  8.69(8)     &   & 3914(2)         & 10.0(4)     &  $-$48(2)  & 0.009(1) & 120(20) & 618    & 53218$-$53778\\
  B1809$-$173& 53105(2)*        &  14.4(5)   &  3(1)        &   & 14.8(6)        & 3.6(5)      &  & 0.27(2)  & 800(100) & 1080   &  52529$-$54394 \\
\vspace{2.1mm} 
             & 54367(13)*       &  1.5(3)    &  0.4(9)      &   & 1.6(4)         & 0.4(9)      &  &  $-$     &   $-$    &  1070  &  53450$-$54764 \\
\vspace{2.1mm} 
  B1815$-$14 & 52057(7)        &  0.54(5)   &  $-$0.7(4)      &   & 0.54(4)        & $-$0.8(4)      &  & $-$      &  $-$     &  511   &  51510$-$52498\\
  B1822$-$09 & 53719--53890    &  7.2(3)    &  $-$15.8(5)  &   &  $-$            &  $-$       &   &  $-$     &   $-$    &        &     \\
\vspace{2.1mm} 
             & 54115.78(4)*    &  122(2)    &  $-$1.8(5)   &   & 121.3(2)        & $-$1.9(3)  &   & $-$      &  $-$     &  935   & 53742$-$54395  \\
 B1823$-$13 & 53238.2(7)*     &  3.4(3)    &  0.07(1)     &   & 3.5(3)         & 0.068(6)    &   & $-$      &  $-$     &  584  & 52518$-$53730  \\
\vspace{2.1mm}  
           & 53734(5)        &  2416(4)   &  13.0(4)     &   & 2417(2)         & 14(5)     &  $-$44(9)  & 0.015(4)  & 75(25)  & 2668   & 53509$-$54080 \\
\vspace{2.1mm} 
 B1838$-$04 & 53408(21)       &  578.7(4)  &  1.7(3)      &   & 578.8(1)        & 1.4(6)     &$-$0.21(1)   & 0.00014(20)  & 80(20)&  340   &  52536$-$54306 \\
\vspace{2.1mm} 
J1853+0545  & 53450(2)*       &  1.49(5)   &  1.6(5)      &   & 1.46(8)         & 3.5(7)     &   & 0.22(5)  & 250(30)  &  333   & 52494$-$54131  \\
\vspace{2.1mm} 
 B1900+06   & 54248(11)*       &  0.33(3)   &  0.04(9)     &   & 0.33(3)         & 0.04(9)    &   &  $-$     &  $-$     &  334   & 52473$-$54821  \\
 B1907+10   & 52700(16)*       &  0.27(7)   &  0.5(7)      &   & 0.27(13)        & 0.5(10)     &   & $-$      & $-$      &  201   & 52470$-$53198 \\
 \vspace{2.1mm} 
           & 53700--54400    &  1.52(5)   &  $-$5.0(7)   &   & $-$             & $-$        &   & $-$      & $-$      &        &      \\
\vspace{2.1mm} 
 B1913+10   & 54162(1)*       &  2.55(3)   &  0.08(5)      &   & 2.52(9)         & 0.3(1)    &   &  $-$     &  $-$     &  294   & 52473$-$54516 \\
 B2224+65   & 54266(14)*        &  0.36(8)   &  $-$0.8(4)   &   & 0.39(7)         & $-$0.6(4)  &   &  $-$     & $-$      &  417   & 53600$-$54706 \\
\hline
\end{tabular}
   \begin{tablenotes}
     \item[] * Glitch epoch determined by phase fit
    \end{tablenotes}
\end{threeparttable}
\end{center}
\end{table}

\end{landscape}

\begin{figure}
\centerline
{
 \hspace{2mm}\psfig{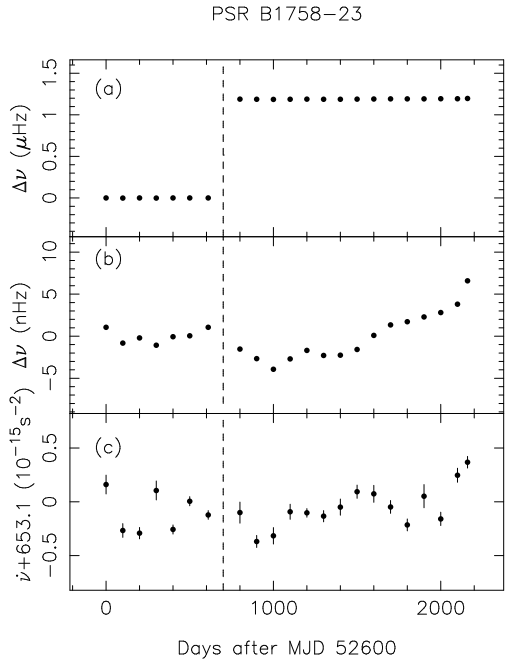}
}
\caption{The glitch of PSR B1758$-$23: (a) variations
of~frequency~ residual~$\Delta\nu$ relative to the pre-glitch
solution, (b) an expanded plot of $\Delta\nu$ where the mean
post-glitch value has been subtracted  from the post-glitch data, and
(c) the frequency first derivative $\dot{\nu}$.}
 \label{Fig:1801n}
\end{figure}

\subsection{PSR B1758$-$23}
This pulsar has a period of 415~ms and large period derivative giving
it a small characteristic age of 58~kyr. PSR B1758$-$23 is distant and
has a large dispersion measure (DM = 1074~cm$^{-3}\,$pc), resulting in a
scattered pulse profile at 1.4 GHz. Furthermore, the pulsar suffers
frequent glitches, so it is difficult to obtain a position measurement
from timing; the value quoted in Table~\ref{tb:position} is from the
VLA observations of \citet{fkv93}.

Six glitches in this pulsar have been reported
\citep{klm+93,sl96,wmp+00,klgj03}, with $\Delta\nu_g/\nu$ in the range
(0.02 -- 0.34)$\times10^{-6}$. Little or no change in $\dot{\nu}$ is
observed, showing that glitches in this pulsar have no significant
post-glitch recovery. Our observations (Fig.~\ref{Fig:1801n}) reveal a
further glitch with $\Delta\nu_g/\nu \sim 0.494\times10^{-6}$ at MJD
$\sim 53309$, the largest seen for this pulsar. As for previous
glitches in this pulsar, there is little or no post glitch decay
(Fig.~\ref{Fig:1801n}b and c), with the fitted value of $Q \sim
0.009$.

\subsection{PSR B1800$-$21}
\begin{figure}
\centerline
{
 \hspace{2mm}\psfig{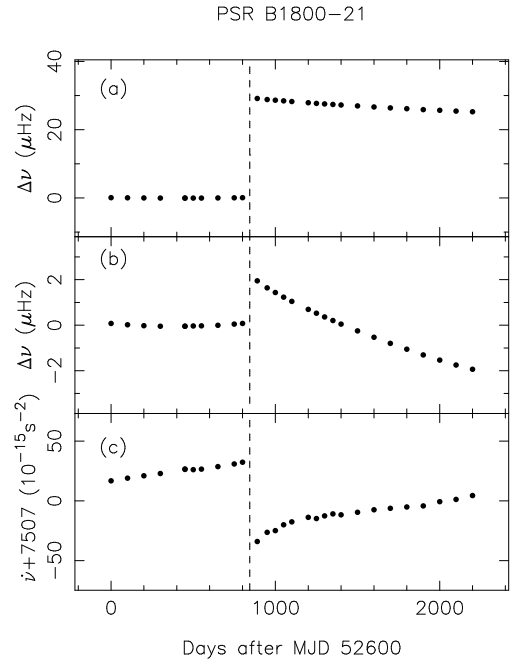}
}
\caption{The glitch of PSR B1800$-$21:  (a) variations of frequency
residual $\Delta\nu$ relative to the pre-glitch solution, (b) an
expanded plot of $\Delta\nu$ where the mean post-glitch value has been
subtracted from the post-glitch data, and (c) the variations of
$\dot{\nu}$.}
\label{Fig:1803n}
\end{figure}

Two giant glitches were detected for PSR B1800$-$21 in 1990 and 1997
with $\Delta\nu_g/\nu \sim 10^{-6}$ \citep{sl96,wmp+00}, while a small
glitch with $\Delta\nu_g/\nu \sim 5\times 10^{-9}$ in 1996 was found
by \citet{klgj03}. Another giant glitch was detected in 2005 March (MJD
$\sim$ 53444) in our observations.  As shown in Fig.~\ref{Fig:1803n},
the fractional jump in frequency is $\Delta\nu_g/\nu \sim
3.9\times10^{-6}$. There is an exponential recovery of a small part of
the change in $\nu$ and $\dot\nu$ with a time constant of about 120
days, but the main observed recovery is linear. A similar linear
increase in $\dot\nu$ following the previous jump can be seen in
Fig.~\ref{Fig:1803n} (c). Once again, this is similar to the Vela
pulsar post glitch behaviour \citep{lpgc96}.  Table~\ref{tb:frequency}
shows a large $\ddot\nu$ both before and after the glitch
corresponding to this linear recovery.

\subsection{PSR B1809$-$173}
\begin{figure}
\centerline
{
 \hspace{2mm}\psfig{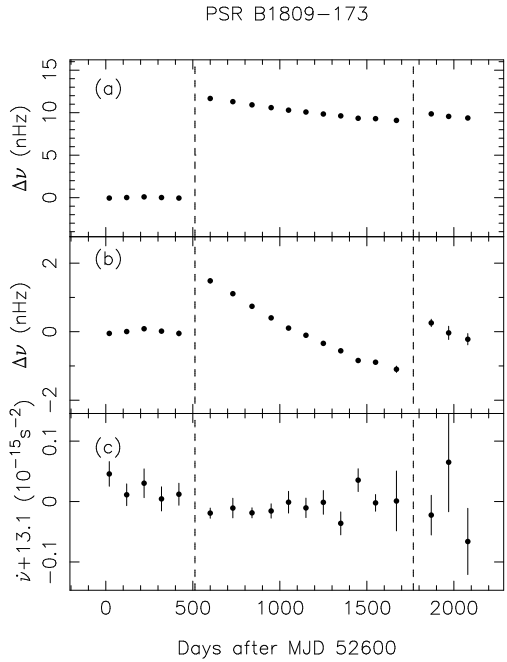}
}
\caption{The glitches of PSR B1809$-$173: (a) variations of frequency
residual $\Delta\nu$ relative to the pre-glitch solution, (b) an
  expanded plot of $\Delta\nu$ where the mean post-glitch value has
  been subtracted  from the post-glitch data, and (c) the frequency first derivative
$\dot{\nu}$.}
\label{Fig:1812n}
\end{figure}

PSR B1809$-$173 suffered a modest jump of fractional size
$\Delta\nu_g/\nu \sim 14.3\times10^{-9}$ and $\Delta\dot\nu/\dot\nu
\sim 3.3\times10^{-3}$ in 2004 April (MJD $\sim$~53105). Observed
frequency variations are shown in Fig.~\ref{Fig:1812n}. An exponential
decay model with $\tau_{\rm d} \sim 800$~d and Q $\sim$ 0.27 is an
adequate representation of the post-glitch relaxation. A second small
glitch at Sep 2007 (MJD $\sim$ 54367) with $\Delta\nu_g/\nu \sim
1.6\times10^{-9}$ is also detected.

\begin{figure}
\centerline
{
 \hspace{2mm}\psfig{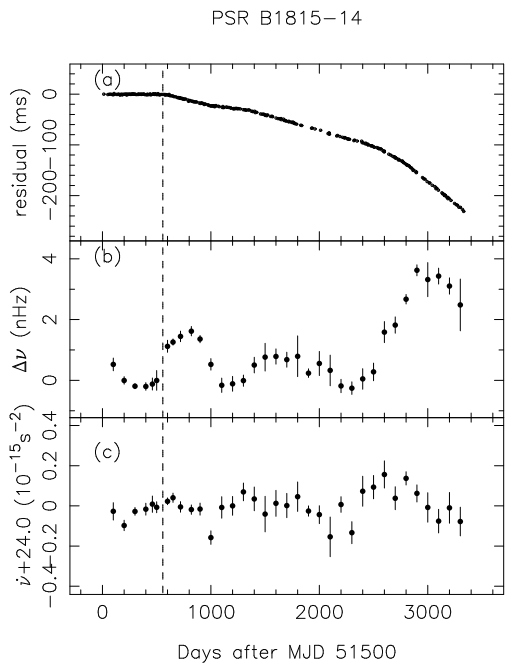}
}
\caption{The glitch of PSR B1815$-$14: (a) timing
residuals  relative to the pre-glitch solution, (b) variations of frequency
residual $\Delta\nu$ relative to the pre-glitch solution, and (c) the variations of
$\dot{\nu}$.}

\label{Fig:1818n}
\end{figure}

\subsection{PSR B1815$-$14}
\begin{figure}
\centerline
{
 \hspace{2mm}\psfig{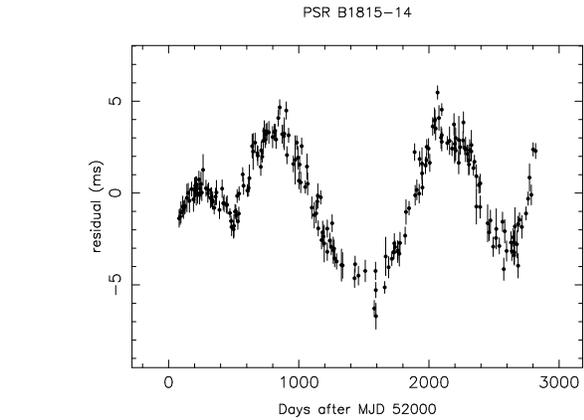}
}
\caption{Timing residuals for PSR B1815$-$14 after the glitch with
  respect to the model given in Table~\ref{tb:frequency}.}
\label{Fig:1818r}
\end{figure}

Fig.~\ref{Fig:1818n} shows that a small glitch was detected in PSR
B1815$-$14 around MJD 52057. Fits of a second-order polynomial to the
pre-glitch data and cubic polynomial to the post-glitch data are given
in Table~\ref{tb:frequency}. Both these and the phase-coherent glitch
fit are consistent with a glitch of magnitude $\Delta\nu_g/\nu \sim
0.54\times10^{-9}$ and $\Delta\dot\nu/\dot\nu \sim -0.8\times
10^{-3}$. The post-glitch data show a quasi-periodic oscillation in
residuals with period $\sim 1200$~d as illustrated in
Fig.~\ref{Fig:1818r}. Similar quasi-periodic oscillations in
  timing residuals have been observed in very long data spans by
  \citet{hlk09}. An increase in $\dot\nu$ (decrease in $|\dot\nu|$)
  was observed from MJD $\sim 54000$ lasting for $\sim 200$~d. This
  event, which is similar to the slow glitches observed in PSR
  B1822$-$09, results in an increase in $\nu$, with $\Delta\nu\sim$~5
  nHz. It is possible that similar but smaller fluctuations in
  $\dot\nu$ at earlier times result in the quasi-periodicity in $\nu$
  and the timing residuals.

\subsection{PSR B1822$-$09}
\begin{figure}
\centerline
{
 \hspace{2mm}\psfig{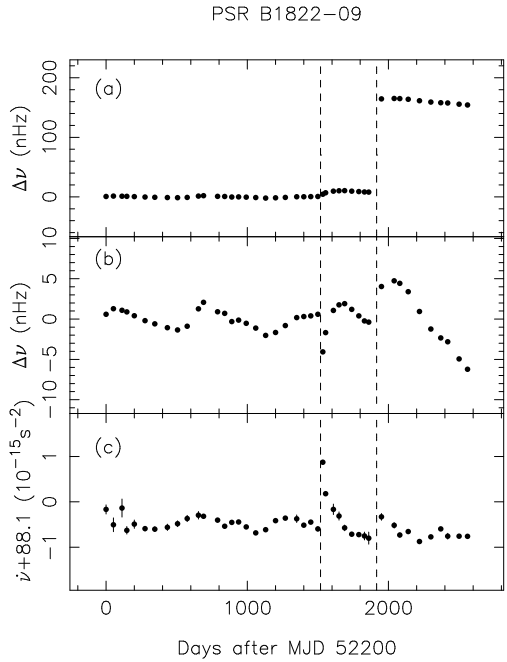}
}
\caption{The glitches of PSR B1822$-$09: (a) variations of frequency
residual $\Delta\nu$ relative to the pre-glitch solution, (b) an
  expanded plot of $\Delta\nu$ where the mean post-glitch value has
  been subtracted  from the post-glitch data, and (c) the variations of $\dot{\nu}$. }
\label{Fig:1825n}
\end{figure}

A total of seven glitches in PSR B1822$-$09 have been reported in the
previous 12 years \citep{sha98,zww+04,sha07}. Five of these glitches
are ``slow'', characterised by a relatively sharp rise in $\dot\nu$
followed by a decline to roughly its pre-glitch state over 100 --
200~d. This results in a gradual spin-up of the pulsar relative to its
pre-glitch state to a new level at which it remains till the next
event. A similar event was observed in PSR J0631+1036 as reported
above in \S~\ref{J0631}. Here we extend the analysis of \citet{zww+04}
by another four years. Fig.~\ref{Fig:1825n} shows that a slow glitch
started at MJD $\sim 53720$ and a ``normal'' glitch occurred at MJD
$\sim$54116 . The former was also seen by \citet{sha07} around MJD
53740 near the end of her data set. The frequency first derivative
$\dot{\nu}$ shows a sudden jump of $ \sim 1.6\times 10^{-15}$ at the
start of the event. The spin frequency increases for about 200~d with
respect to the pre-glitch fit until the increase in $\dot\nu$ decays
away.

For the larger and latest glitch, $\Delta\nu_g/\nu \sim
0.121\times10^{-6}$. Probably because of timing noise and the short
post-glitch data span, no clear post-glitch  exponential recovery can be seen.
\citet{zww+04} suggested that a small-magnitude slow glitch probably
occurred near the end of their data span around MJD 52900. Although
there is a small fluctuation in $\nu$ and $\dot\nu$ visible in
Fig.~\ref{Fig:1825n} at this time, we consider that it is not clear
enough to be identified as a slow glitch.

\subsection{PSR B1823$-$13}
\begin{figure}
\centerline
{
 \hspace{2mm}\psfig{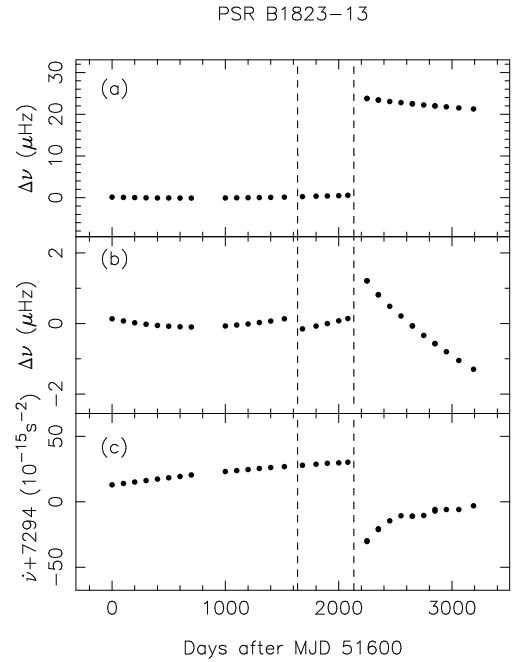}
}
\caption{The glitches of PSR B1823$-$13: (a) variations of frequency
residual $\Delta\nu$ relative to the pre-glitch solution, (b) an
expanded plot of $\Delta\nu$ where the mean post-glitch value has been
subtracted from the post-glitch data, and (c) the frequency first
derivative $\dot{\nu}$.}
\label{Fig:1826n}
\end{figure}

PSR B1823$-$13 is a young pulsar ($\sim$ 21.4~kyr) with a period of
101.4 ms. Two large glitches have been previously reported by
\citet{sl96} with $\Delta\nu_g/\nu \sim 2.7\times10^{-6}$ and
$\Delta\nu_g/\nu \sim 3.0\times10^{-6}$ in 1986 and 1992
respectively. Both had a partial exponential recovery with timescale
$\sim 100$~d followed by a linear increase in $\dot\nu$. As shown in
Fig.~\ref{Fig:1826n}, two new glitches were detected in the Nanshan
timing. A small glitch with $\Delta\nu_g/\nu \sim 3.5\times10^{-9}$
occurred at MJD $\sim$ 53238, and a large glitch at MJD $\sim$ 53734
with $\Delta\nu_g/\nu \sim2.4 \times10^{-6}$. There is no obvious
relaxation after the small glitch, but after the large glitch the
relaxation is very similar to that of the earlier large glitches in
this pulsar. Fitting for a single exponential decay model with
$\tau_{d} \sim$ 75~d and $Q\sim$ 0.015 gives a relatively good fit to
the post-glitch data between MJD 53509 and 54080. Because of the small
$Q$, there is effectively a permanent change in $\dot\nu$,
$\Delta\dot\nu_p \sim -44\times10^{-15}$~s$^{-2}$.  Fig.~\ref{Fig:1826n}
shows a linear increase in $\dot\nu$ both before (through the small
glitch) and after the larger glitch. Fitting to the linear part of the
pre- and post-glitch data, despite the short data span after
the glitch, we have $\ddot\nu$ approximately  $100 \times 10^{-24}$~s$^{-3}$ and
 $65\times 10^{-24}$~s$^{-3}$ respectively. We will further discuss this in
\S\ref{sec:dis}.

\subsection{PSR B1838$-$04}

\begin{figure}
\centerline
{
 \hspace{2mm}\psfig{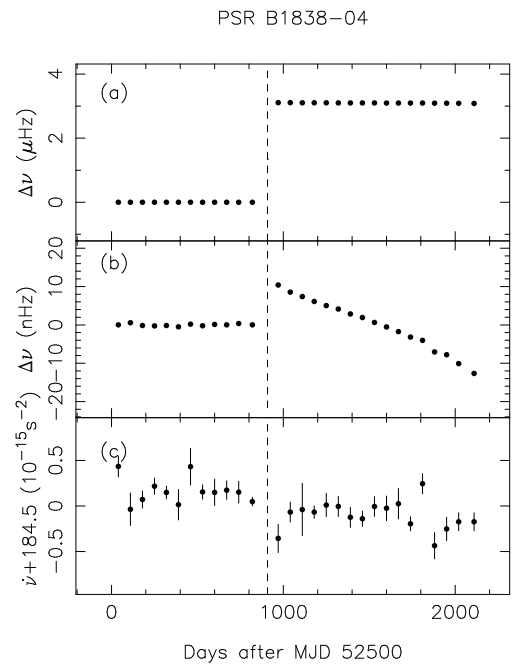}
}
\caption{The glitch of PSR B1838$-$04: (a) variations of frequency
residual $\Delta\nu$ relative to the pre-glitch solution, (b) an
  expanded plot of $\Delta\nu$ where the mean post-glitch value has
  been subtracted  from the post-glitch data,  and (c) the variations of $\dot{\nu}$.}
\label{Fig:1841n}
\end{figure}

A large glitch with $\Delta\nu_g/\nu \sim 0.58\times10^{-6}$ was
observed in PSR B1838$-$04 between MJD 53387 and 53429. This is the
first glitch observed in this middle-aged pulsar which was discovered
by \citet{cl86}.  Fig.~\ref{Fig:1841n} shows a hint of a short
exponential decay, but the main effect of the glitch is what seems
to be a permanent change in $\dot\nu$. Fitting for this with a 80~d
exponential decay gave a value of $\Delta\dot\nu_p $ of
$-0.21\times10^{-15}~$s$^{-2}$ (Table~\ref{tb:glitch}).

\begin{figure}
\centerline
{
 \hspace{2mm}\psfig{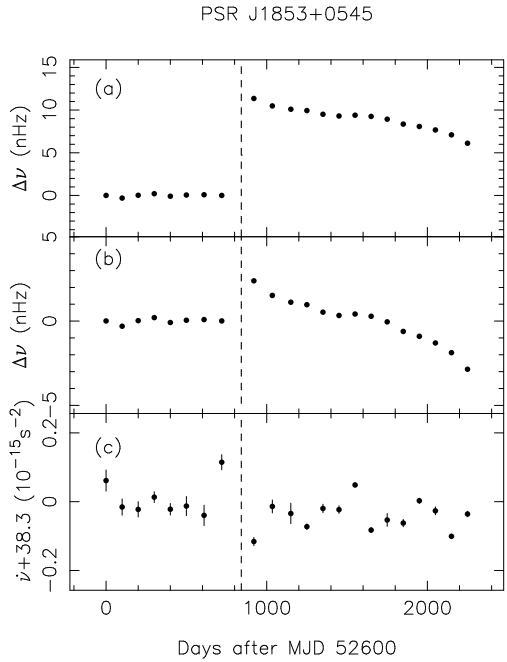}
}
\caption{The glitch of PSR J1853+0545: (a) variations of frequency
residual $\Delta\nu$ relative to the pre-glitch solution, (b) an
expanded plot of $\Delta\nu$ where the mean post-glitch value has been
subtracted from the post-glitch data, and (c) the frequency first
derivative $\dot{\nu}$.}
\label{Fig:1853n}
\end{figure}

\subsection{PSR J1853+0545}
We have detected the first known glitch for this recently discovered
\citep{kbm+03} and relatively old pulsar (3.3 Myr). The
glitch is small with a magnitude of $\Delta\nu_g/\nu \sim
1.46\times10^{-9}$.
Fig.~\ref{Fig:1853n} shows a partial
exponential recovery; fitting for this gives $\tau_{\rm d} \sim
250$~d, and fractional decay $Q \sim 0.22$.

\begin{figure}
\centerline
{ \hspace{2mm}\psfig{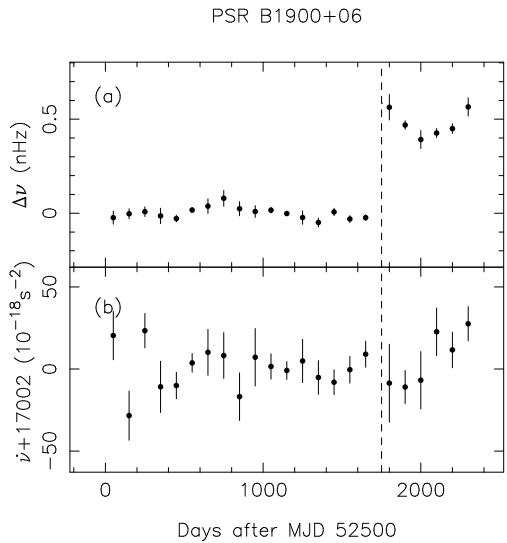}
}
\caption{The glitch of PSR B1900+06:  (a) variations of frequency
residual $\Delta\nu$ relative to the pre-glitch solution, (b) the frequency first
derivative $\dot{\nu}$.}
\label{Fig:1900n}
\end{figure}

\subsection{PSR B1900+06}

This pulsar has been monitored at Nanshan since 2002 and a small
glitch was detected in May 2007 (MJD $\sim$ 54248). The fractional
size is $\Delta\nu_g/\nu \sim 0.33\times10^{-9}$.
Fig.~\ref{Fig:1900n} shows that where may be some post-glitch
recovery, but the data span is too short for further analysis.

\begin{figure}
\centerline
{
\hspace{2mm}\psfig{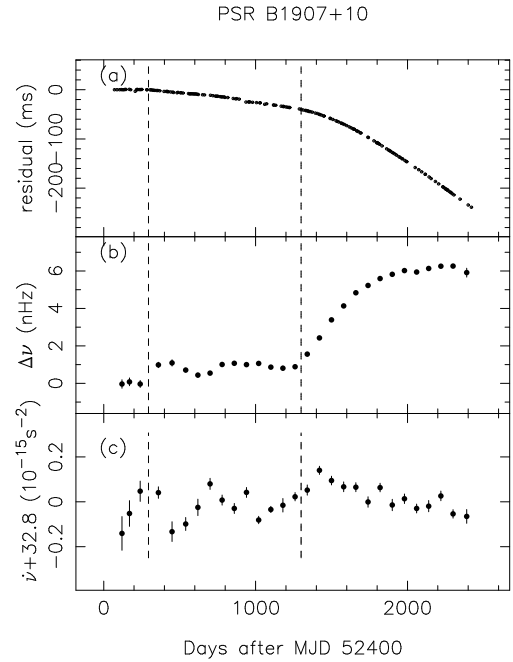}
}
\caption{The glitches of PSR B1907+10: (a) timing residuals relative to
the pre-glitch solution, (b) frequency residual $\Delta\nu$ relative
to the pre-glitch solution, and (c) the variations of $\dot{\nu}$.}
\label{Fig:1909n}
\end{figure}

\subsection{PSR B1907+10}
PSR B1907+10 has a period of 284~ms and is relatively old ($\tau_c
\sim 1.7$~Myr).  Fig.~\ref{Fig:1909n} shows a tiny glitch at MJD
$\sim$ 52700 with $\Delta\nu_g/\nu \sim 0.27\times10^{-9}$. Following
that, there is evidence for a slow glitch between MJD  $\sim$ 53700 and
 $\sim$ 54400. A gradual but significant increase in frequency of about 5~nHz
was observed
with a corresponding perturbation in $\dot\nu$. Although this may be
just a fluctuation in the timing noise, it does have the
characteristics of the slow jumps observed in PSR B1822$-$09 and some
other pulsars.

\begin{figure}
\centerline
{
\hspace{0mm}\psfig{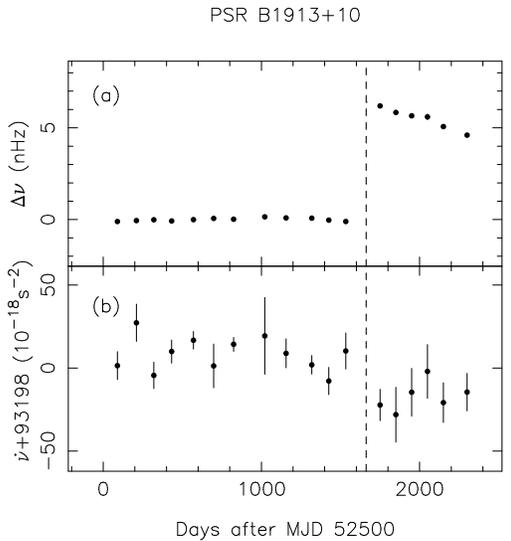}
}
\caption{The glitch of PSR B1913+10: variations of (a) frequency
residual $\Delta\nu$ relative to the pre-glitch solution, (b) the
frequency first derivative $\dot{\nu}$.}
 \label{Fig:1915n}
\end{figure}

\subsection{PSR B1913+10}
This pulsar has a period of 405~ms and characteristic age $\tau_c \sim
0.4$~Myr. Fig.~\ref{Fig:1915n} shows that a small glitch with
$\Delta\nu_g/\nu \sim 2.5\times10^{-9}$ was observed at MJD $\sim$
54162. Although the post-glitch data span is small, there is clear
evidence for an increase in $|\dot\nu|$ indicating some post-glitch recovery.

\begin{figure}
\centerline
{
\hspace{0mm}\psfig{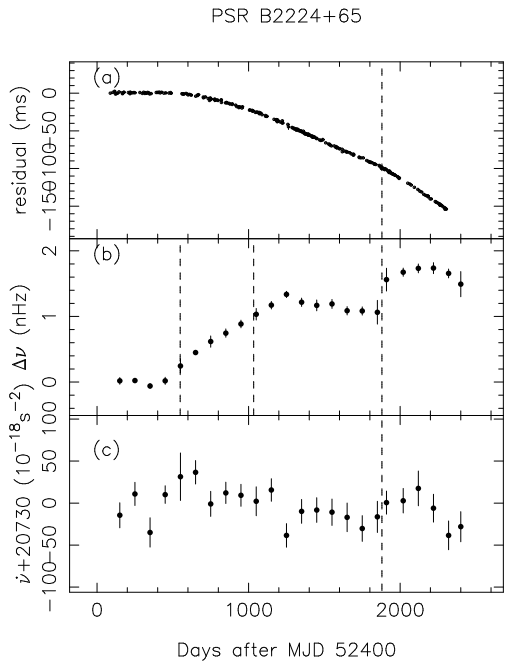}
}
\caption{Glitches of PSR B2224+65: (a) timing residuals relative to
pre-glitch model, (b) variations of frequency residual $\Delta\nu$
relative to the pre-glitch solution, and (c) the variations of
$\dot{\nu}$. The short dashed lines in the $\Delta\nu$ plot mark the
small glitches claimed by \citet{js06} and the longer dashed line
marks the glitch detected in our data.}
 \label{Fig:2225n}
\end{figure}

\subsection{PSR B2224+65}
PSR B2224+65 is the high-velocity pulsar associated with the Guitar
Nebula \citep{crl93}. It has a period of 0.682~s and a characteristic
age of 1.1~Myr. A giant glitch with $\Delta\nu_g/\nu \sim
1.7\times10^{-6}$ occurred in 1976 October \citep{btd82} and three
tiny glitches ($\Delta\nu_g/\nu < 0.2\times10^{-9}$) between 2000
December and 2005 March were reported by \citet{js06}. 
The long dashed line in Fig.~\ref{Fig:2225n} marks another
  glitch detected in the Nanshan data. This event at MJD $\sim$ 54266
has $\Delta\nu_g/\nu \sim 0.39\times10^{-9}$. There is some evidence
that $|\dot\nu|$ decreased after the glitch, similar to PSR B0144+59
(\S~\ref{B0144}). 

The short dashed lines mark the second and third small
  glitches claimed by \citet{js06} at MJD 52950 and 53434 for which
  the estimated $\Delta\nu_g$ values were $0.12\times 10^{-9}$~Hz and
  $0.28\times 10^{-9}$~Hz, respectively. (Nanshan observations
  commenced in 2002 August, after the time of the first small glitch.)
  Fig.~\ref{Fig:2225n}(b) shows that the pulse frequency was
  increasing (relative to the pre-glitch solution) across the interval
  covered by these claimed glitches, but the increase appears smooth
  and no glitch larger than about $0.1\times 10^{-9}$~Hz is evident in
  the data. As for PSR B0525+21, \citet{js06} did not give phase or
  frequency residual plots for this pulsar, so it is not possible to
  assess the evidence for their claimed glitches.

\section{DISCUSSION} \label{sec:dis}
The study of pulsar glitches provides one of the few ways of probing
the interior of neutron stars. Previous studies have shown that
glitch properties, specifically characteristics of the post-glitch
recovery, vary greatly from one pulsar to another and for different
glitches in the same pulsar. Furthermore, while inter-glitch intervals
are relatively short ($ \lapp  1$~yr) in some young pulsars, they are
much longer in older pulsars, with many having no observed glitch over
observing spans of 30 years or more. It is therefore valuable to
continue timing observations of a large sample of pulsars in order to
better characterise glitch properties and hopefully to lead to a better
understanding of the mechanisms involved and the underlying properties
of neutron stars.

\begin{figure}
\centerline
{
\hspace{0mm}\psfig{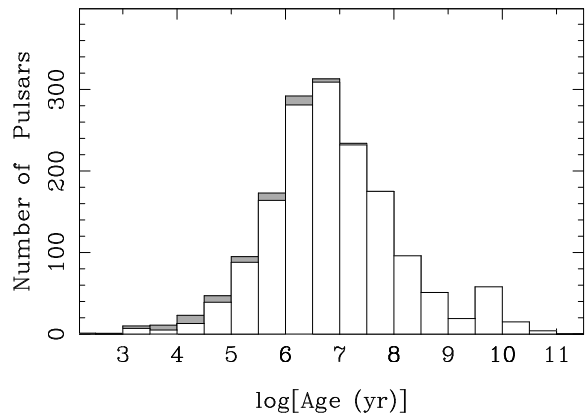}
}
\caption{Distribution of known pulsars
versus pulsar characteristic age with the distribution of glitching
pulsars marked by the shaded area.}
\label{Fig:histo_age}
\end{figure}

This paper reports on glitches detected in continuing observations of
a sample of 280 pulsars using the 25-m Nanshan telescope of Urumqi
Observatory. We detected 29 glitches in 19 pulsars, with this being
the first detection of a glitch in 12 of these pulsars. With those
already listed in the ATNF Pulsar Catalogue, this increases
the number of pulsars with detected glitches to 63 and the total
number of detected glitches to 199. Fig.~\ref{Fig:histo_age} shows the
distribution of glitching pulsars among all known pulsars as a
function of characteristic age. About half of all known pulsars with
ages less than $3\times 10^4$~yr have detected glitches, but for older
pulsars the fraction is much less, with no glitches being detected in
pulsars older than $3\times 10^7$~yr.

\begin{figure}
\centerline
{
\hspace{0mm}\psfig{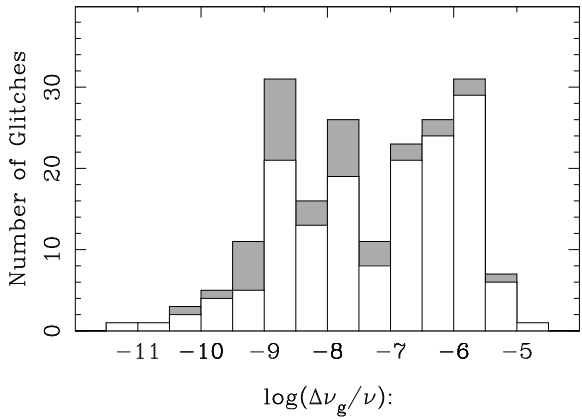}%
}
\caption{Distribution of relative glitch amplitudes of all known
  glitches. Those detected at Urumqi Observatory are marked
 by the shaded  area. }
\label{Fig:histo_size}
\end{figure}

The range of fractional glitch amplitude $\Delta\nu_g/\nu$ for
glitches detected in our work is from $4.4\times10^{-11}$ to
$3.9\times10^{-6}$. Fig.~\ref{Fig:histo_size} shows the distribution
of fractional glitch amplitude for all known glitches\footnote{The
  three small glitches claimed by \citet{js06} for which we see no
  evidence have been omitted from this and subsequent
  figures.}. The detection threshold for a glitch depends upon
  the quality of the TOAs, the frequency of the observations and the
  intrinsic timing noise. TOA uncertainties depend on the telescope
  and receiver parameters and the pulsar flux density and pulse
  width. For the Nanshan observations, they are typically 100 -- 300
  $\mu$s. The pulsar timing project at Nanshan benefits from frequent
  and regular observations on a large sample of pulsars. Although
  there are some longer gaps due to instrumental or scheduling issues,
  observations of all pulsars are typically at 10-day intervals. The
  level of intrinsic quasi-continuous timing noise varies greatly from
  pulsar to pulsar. This and the overall sensitivity of the
  observations to small glitches is best assessed from the $\Delta\nu$
  plots given for each pulsar. Typically, the sensitivity is in the
  range $(0.1 - 1.0)\times 10^{-9}$~Hz. Our work has revealed eight
small glitches with relative size $<10^{-9}$. These studies have
extended the sample of small glitches and indicate that the frequency
of occurrence of small glitches is comparable to that of ``giant''
glitches with $\Delta\nu_g/\nu  \gapp 10^{-6}$ \citep[cf.,][]{lsg00}.

\begin{figure}
\centerline
{
\hspace{0mm}\psfig{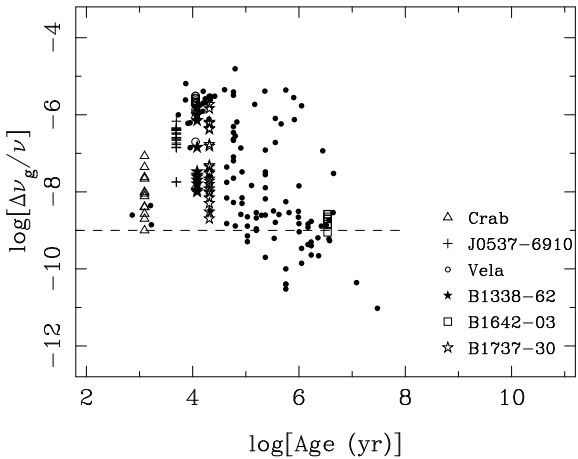}
}
\caption{Fractional glitch amplitude $\Delta\nu_{g}/\nu$ versus
characteristic age. The dash line marks the amplitude of
$\Delta\nu_g/\nu$ = 10 $^{-9}$.~}
\label{Fig:size_age}
\end{figure}

Fig.~\ref{Fig:size_age} shows glitch fractional amplitudes plotted
against the pulsar age. Although there is large scatter, there is a
clear tendency for the average $\Delta\nu_g/\nu$ to increase with age
up to somewhere between $10^4$ and $10^5$ years and then to decrease
with increasing age. The increased detection of small glitches
reported in this and other recent papers
\citep{klgj03,cb04,js06} has clarified this trend. Very small
glitches with $\Delta\nu_g/\nu < 10^{-9}$ are only seen in pulsars
with age greater than $10^5$~yr.

\citet{lsg00} found that the jump in $\dot\nu$ at the time of a
glitch, $\Delta\dot\nu_g$, was related to the value of $\dot\nu$, with
the fractional change $\Delta\dot\nu_g/\dot\nu$ being approximately
constant. Table~\ref{tb:glitch} shows that most of the measured values
are about $10^{-3}$, but there is a scatter of up to two orders of
magnitude in the ratio, even for different glitches in a given
pulsar.

\begin{figure}
\centerline
{
\hspace{0mm}\psfig{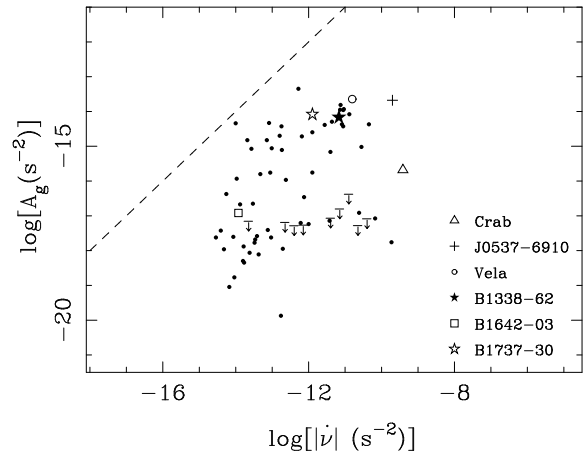}
}
\caption{Plot of glitch activity parameter, A$_{g}$ versus the
absolute value of the spin-down rate $\dot{\nu}$. The dash indicates
equal values of these two quantities. Upper limits of $A_g$ are
given by assuming one glitch $\Delta\nu_g/\nu = 10^{-9}$ occurred for
young pulsars with no detected glitch and $\tau_{c} <$ 10$^5$~yr.
Frequent glitching pulsars are marked specifically.}
\label{Fig:agvsnudot}
\end{figure}

The glitch activity parameter, $A_g$ (Equation~\ref{eq:activity}) is a
way of quantifying the contribution of glitches to $\dot\nu$. Since
pulsars slow down, $\dot\nu$ is negative, but glitches are spin-ups
and hence the contribution is positive. In some models, the fractional
contribution is related to the fraction of the neutron-star moment of
inertia which is in superfluid form
\citep[e.g.,][]{rzc98}. Fig.~\ref{Fig:agvsnudot} shows the dependence
of activity parameter on $|\dot\nu|$. Again there is a large scatter,
but there is a tendency for increasing activity with increasing
$|\dot\nu|$ with the ratio between them ranging between $10^{-2}$ and
$10^{-5}$. There seems to be a turn-over for the pulsars with largest
$|\dot\nu|$, for example the Crab pulsar and PSR J0537$-$6910, having
relatively smaller activity parameters.  There is one pulsar, PSR
J2301+5852 (the 7~s anomalous X-ray pulsar 1E~2259+586 in the
supernova remnant CTB 109), where this ratio is close to 1.0. However,
it should be noted that only one glitch has been observed in this
pulsar \citep{kgw+03}, so the derived activity parameter is
effectively an upper limit.

Glitches are generally attributed to the release of stress built up as
a result of the steady spin-down of the pulsar. This stress may be in
the solid crust, relieved by crustal starquakes \citep{bppr69}, or
may be on pinned vortices in the superfluid interior, relieved by a
collective unpinning of many vortices \citep{ai75}. Both of these
result in a sudden spin-up of the crust and hence a jump in the
observed pulse frequency. In these models, one might expect a
correlation of glitch size with the duration of the preceding or
following inter-glitch interval, depending on the detailed
mechanism. A strong correlation between the
size of glitches and the length of the following inter-glitch interval
was observed for PSRs J0537$-$6910  \citep{mmw+06} and B1642$-$03
\citep{sha09}, but such a clear correlation is not observed in other
pulsars.

\begin{figure}
\centerline
{
\hspace{0mm}\psfig{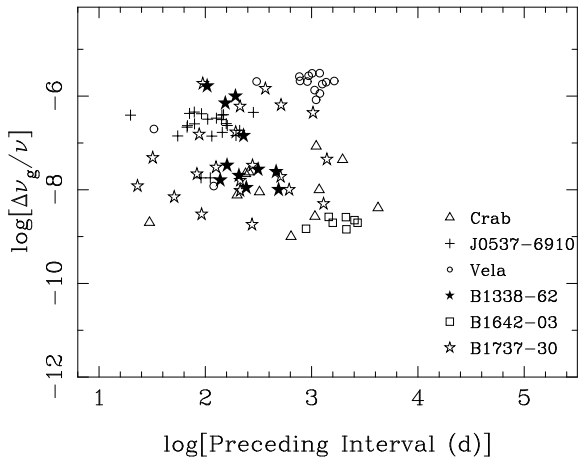}
}
\caption{Fractional glitch amplitude versus length of the
preceding interglitch interval.}
\label{Fig:size_intervalP}
\end{figure}

\begin{figure}
\centerline
{
\hspace{0mm}\psfig{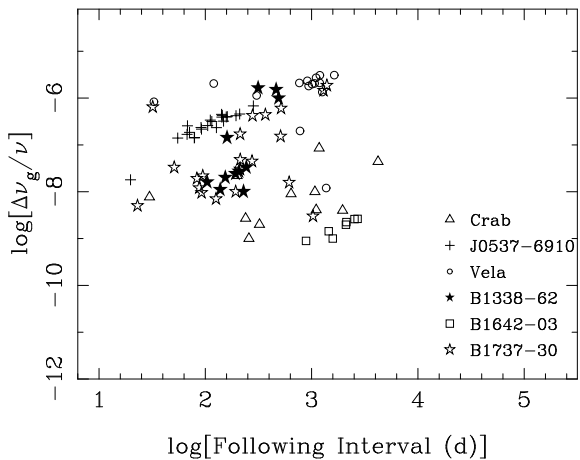}
}
\caption{Fractional glitch amplitude versus length of the
following interglitch interval.}
\label{Fig:size_intervalF}
\end{figure}

Fig.~\ref{Fig:size_intervalP} and ~\ref{Fig:size_intervalF} shows
the dependence of the fractional glitch amplitude on the length of
the preceding and following interglitch intervals respectively for
the most frequently glitching pulsars. Despite the naive expectation
that glitches would be large after a long build-up period, there is
very little correlation of glitch amplitude with length of the
preceding interval. For example, the Crab
pulsar and PSR B1737$-$30 have a wide range of both parameters and there is
essentially no correlation.  Most glitches in the Vela pulsar are
large and have long inter-glitch intervals. Two smaller glitches
have been observed with shorter preceding intervals, but the
preceding interval for the smallest glitch is more than four times
the interval for the next smallest which is more than an order of
magnitude larger. For PSR J0537$-$6910, there is no correlation and,
for PSR B1338$-$62, if anything there is a negative
correlation. For post-glitch intervals,
Fig.~\ref{Fig:size_intervalF} shows the correlations found by
\citet{mmw+06} and \citet{sha09} for PSR J0537$-$6910 and PSR
B1642$-$03 respectively, and a possible steeper positive dependence
for PSR B1338$-$62. Again for the Crab and Vela pulsars and
PSR B1737$-$30 there is little correlation.

\begin{table*}
\centering
\begin{minipage}{150mm}
\caption{Rate of change of $\dot\nu$ for pulsars which show a linear
  post-glitch recovery.}
\label{tb:lnuddot}
\begin{tabular}{@{}lrrlllll@{}}
\hline
Pulsar     &  P~   & Age        & Max $\Delta\nu_{g}/\nu$  & $\ddot{\nu}$            & $n$        & Note \\
 Name      & (s)  & (kyr)       & ($10^{-6}$)              & ($10^{-24}$~s$^{-3}$)   &          &  \\
\hline
B0833$-$45  & 0.089  &  11.3   & 3.06(7)    & 873(10)   & 40.3(5)     & MJD 50024--50364 \citep{wmp+00}  \\
\\                                                                  
B1046$-$58  & 0.124  &  20.3   & 2.995(7)   &  72(10)   & 15(2)       & MJD 47909--48929 \citep{wmp+00}  \\
\\                                                                  
B1706$-$44  & 0.102  &  17.5   & 2.057(2)   &  124(1)   & 15.42(1)    & MJD 47909--48746 \citep{wmp+00}  \\
\\                                                                  
B1727$-$33  & 0.139  &  26     & 3.033(8 )  & 100.5(13) &  37.6(5)    & MJD 48100--49500 \citep{sl96}\\
\\                                                                  
B1737$-$30  & 0.606  &  20.6   & 1.8536(14) & 12.69(1)  &  13.07(1)   & MJD 53250--54394 (this work) \\
\\                                                                  
B1800$-$21  & 0.133  &  15.8   & 4.073(16)  & 236.5(1)  &  31.77(1)   & MJD 51916--53429 (this work) \\
            &        &         &            & 188.1(1)  &  24.89(1)   & MJD 53660--54832 (this work)\\
            &        &         &            & 107(30)   &  14(4)      & MJD 46600--48425 \citep{sl96}\\
\\                                                                  
B1823$-$13  & 0.101 &  21.4    & 3.05(5)    & 106.3(5)  &  19.74(1)   & MJD 51574--53199 (this work) \\
            &       &          &            &   65(2)   &  12.3(2)    & MJD 53253--53723 (this work) \\
            &       &          &            & 203.0(1)  &  37.53(4)   & MJD 53950--54832 (this work)  \\
\hline
\end{tabular}
\end{minipage}
\end{table*}

Pulsars have a wide variety of post-glitch behaviours. Most show some
form of relaxation toward the extrapolated pre-glitch solution after a
glitch. However, there are some cases where the pulsar simply keeps
spinning at the new rate with no change in $\dot\nu$. In terms of the
exponential decay model (Equations \ref{eq:glitch} --
\ref{eq:glitch5}), $\Delta\nu_d \approx 0$, $\Delta\nu_g \approx
\Delta\nu_p$, $Q \approx 0$ and $\Delta\dot\nu_g \approx 0$. PSR
B1758$-$23 is a good example of a pulsar with large glitches having
these characteristics. For small glitches, it is generally not
possible to measure any decay since it is masked by random and
systematic timing noise.

In some pulsars there is an apparently permanent change in $\dot\nu$
at the time of a glitch, for example, PSRs B0144+59, B0402+61,
B1838$-$04 and B2224+65. For PSR B0144+59 and possibly PSR B2224+65,
the change is positive which is unusual and not in accord with the
exponential decay model. For the others, it is of course possible that
the decay is exponential but with a time constant much longer than our
post-glitch data span.

Exponential recoveries are seen in many pulsars, but these generally
have relatively short timescales, $\sim 100$~d, and low $Q$, that is,
only a fraction of the initial frequency jump decays. Examples are the
first observed glitch in PSR J0631+1036 and the large glitches in PSRs
B1800$-$21 and B1823$-$13. For PSR B1838$-$04, there is a short
exponential recovery preceding the long-term decay mentioned above.

These small exponential recoveries are often associated with other
post-glitch behaviours. For example, they precede linear increases in
$\dot\nu$ in PSRs B1737$-$30, B1800$-$21 and B1823$-$13. The latter
two pulsars also show an effectively permanent increase in $|\dot\nu|$
at the time of the glitch, that is, the linear post-glitch recovery is
approximately parallel to but below that existing before the observed
glitch (see Figures \ref{Fig:1803n} and \ref{Fig:1826n}). Very
similar behaviour is observed in the Vela pulsar \citep{lpgc96}, and in
PSRs B1046$-$58, B1706$-$44 and B1727$-$33
\citep{sl96,wmp+00}. Table \ref{tb:lnuddot} gives the observed
$\dot\nu$ slopes, i.e., $\ddot\nu$, for the linear portion of the
recovery. The slope in the Vela pulsar is an order of magnitude
greater than for the other pulsars. For PSRs B1800$-$21 and
B1823$-$13, multiple glitches with linear behaviour are observed, but
the slopes are distinctly different for the different glitches,
varying by about a factor of two. Similar slope variations are evident
in the Vela data \citep{lpgc96}. The apparent braking indices
\begin{equation}
n = \frac{\ddot\nu \nu}{\dot\nu^2}
\label{eq:brake}
\end{equation}
corresponding to the linear increases in $\dot\nu$ are also given in
Table \ref{tb:lnuddot}. Values range between 12 and 40, emphasising
that these changes are due to internal dynamics in the neutron star,
not to magnetic-dipole braking (for which $n=3$).  Although the Vela
pulsar has the largest values, there is no clear correlation
of either the slope or the braking index with pulse period or
characteristic age. Table~\ref{tb:lnuddot} also gives the
  maximum observed fractional glitch size, showing that all of these
  pulsars have giant glitches. However, it is worth noting that linear
  recoveries are not observed in PSRs B1338$-$62, J1617$-$5055,
  B1757$-$24 and J2021+3651, despite these having similar
  characteristic ages and similar maximum glitch sizes.

\begin{figure}
\centerline
{
\hspace{0mm}\psfig{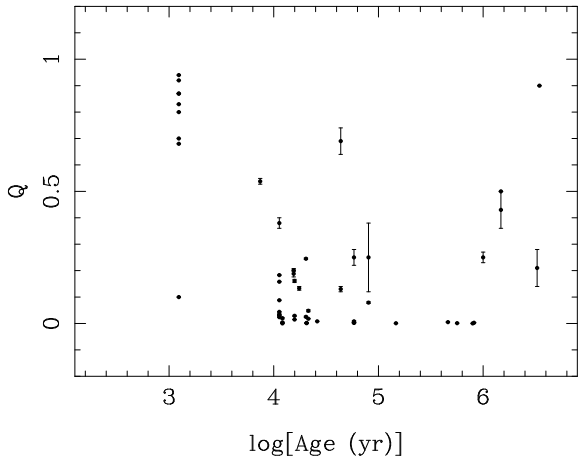}
}
\caption{The fraction parameter Q in which a glitch $\Delta\nu_g$
decays, versus characteristic age.}
\label{Fig:qvsage}
\end{figure}

\begin{figure}
\centerline
{
\hspace{0mm}\psfig{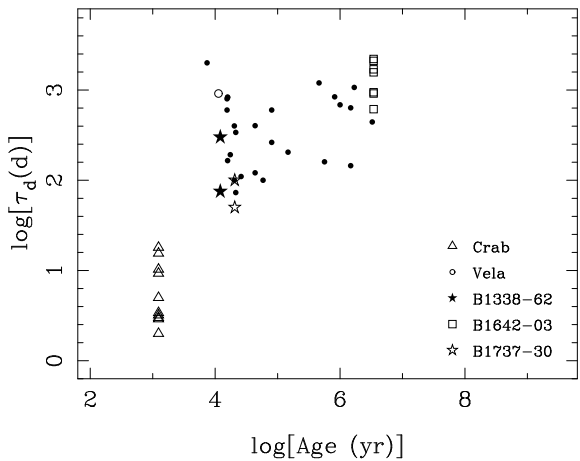}
}
\caption{Glitch decay time-scale versus pulsar
characteristic age. }
\label{Fig:taud_age}
\end{figure}

Fig. ~\ref{Fig:qvsage} shows the fraction of $\Delta\nu_g$ which
decays, $Q$, versus characteristic age. Except for one event with $Q \sim
0.1$, the Crab pulsar usually has a large $Q$ between 0.7 and 1.0.
Although there is some tendency for decreasing $Q$ with increasing age
\citep{wmp+00}, several old pulsars, for example, PSRs B0525+21,
 B1642$-$03, B1809$-$173 and J1853+0545 show some evidence for exponential
 relaxation. Fig. ~\ref{Fig:taud_age} shows that, with the addition of PSR 
B1642$-$03 \citep{sha09}, there is a clearer correlation of decay
timescale with age than was previously evident \citep{wmp+00}. This
suggests a connection with the internal temperature of the neutron
star \citep[cf.][]{acp89}.

Two new examples of the ``slow glitch'' phenomenon, in PSRs J0631+1036,
and B1907+10, have been identified in this paper. Previously,
slow glitches have only been identified in one pulsar, PSR B1822$-$09
\citep{sha98,su00,zww+04,sha05,sha07}. Although it is possible to
interpret these events as fluctuations in timing noise, they appear to
be distinct and repeatable with similar characteristics. Hence it is
reasonable to consider them as a separate category. PSR B1822$-$09 has
had five slow glitches between 1995 and 2004 and one more in 2006
reported in this work.  The new
slow glitches detected in PSRs J0631+1036 and B1907+10 significantly
increase the sample of slow glitches. The event in PSR J0631+1036 is
similar in size and characteristics to those in PSR B1822$-$09, with an
increment in $\dot\nu$ of about $2\times 10^{-15}$~s$^{-2}$ lasting
$\sim 100$~d. However it differs in that the decline in $\dot\nu$ overshot
the pre-event value, resulting in an effective spin-down of the period
to approximately its extrapolated value. In PSR B1822$-$09, the
spin-ups persist until the next event.

The results presented in this paper significantly increase the sample
of known pulsar glitches. They further illustrate the complex nature
of glitches, with a great diversity of glitch sizes and modes of
post-glitch recovery being observed, both in individual pulsars and
across different pulsars. They also strengthen the identification of
slow glitches as a distinct phenomenon. While there are several
different models for glitches, as yet none of these models
satisfactorily account for this diversity and so the promise that
glitches have as probes of the interior of neutron stars has not yet been
fully realised. Hopefully, these and future results will act as a
stimulus for a better understanding of the mechanisms involved and
their implications for the physics of neutron-star interiors.

\subsection*{ACKNOWLEDGMENTS}
This work was supported by the Knowledge Innovation Program of CAS,
Grant No. KJCX2-YW-T09, NSFC project No. 10673021, and National Basic
Research Program of China (973 program 2009CB824800).


\label{lastpage}

\end{document}